\documentclass[a4paper,fleqn,usenatbib]{mnras}
\usepackage{amsmath}
\usepackage{mathptmx}
\usepackage{txfonts}
\usepackage[T1]{fontenc}
\usepackage{graphicx}	
\usepackage{amssymb}	
\usepackage[normalem]{ulem}
\usepackage{longtable}
\usepackage{multirow}

\newcommand{\be}{\begin{equation}}
\newcommand{\e}{\end{equation}}
\newcommand{\bear}{\begin{eqnarray}}
\newcommand{\ear}{\end{eqnarray}}

\newcommand{\hmpc}{{\, h^{-1}\, {\rm Mpc}}}

\def\apj{ApJ}
\def\apjl{ApJL}
\def\apjs{ApJS}

\def\mnras{MNRAS}
\def\aap{A\&A}

\def\nat{Nature}
\def\apjs{ApJS}
\def\apjl{ApJ Letters}

\title[Significance of Mutual Information]{A study on the statistical
  significance of mutual information between morphology of a galaxy
  and its large-scale environment}

\author[Sarkar, S. and Pandey, B.] {Suman
  Sarkar$^{}$\thanks{suman2reach@gmail.com} and Biswajit
  Pandey$^{}$\thanks{biswap@visva-bharati.ac.in} \\$^1$ Department of
  Physics, Visva-Bharati University, Santiniketan, Birbhum, 731235,
  India }

 \date{\today}

 \pubyear{2020}
  
\begin{document}
\label{firstpage}
\pagerange{\pageref{firstpage}--\pageref{lastpage}}      
\maketitle
       
\begin{abstract}
  A non-zero mutual information between morphology of a galaxy and its
  large-scale environment is known to exist in SDSS upto a few tens of
  Mpc. It is important to test the statistical significance of these
  mutual information if any. We propose three different methods to
  test the statistical significance of these non-zero mutual
  information and apply them to SDSS and Millennium Run simulation. We
  randomize the morphological information of SDSS galaxies without
  affecting their spatial distribution and compare the mutual
  information in the original and randomized datasets. We also divide
  the galaxy distribution into smaller subcubes and randomly shuffle
  them many times keeping the morphological information of galaxies
  intact. We compare the mutual information in the original SDSS data
  and its shuffled realizations for different shuffling lengths. Using
  a $t$-test, we find that a small but statistically
  significant (at $99.9\%$ confidence level) mutual information
  between morphology and environment exists upto the entire length
  scale probed. We also conduct another experiment using mock datasets
  from a semi-analytic galaxy catalogue where we assign morphology to
  galaxies in a controlled manner based on the density at their
  locations. The experiment clearly demonstrates that mutual
  information can effectively capture the physical correlations
  between morphology and environment. Our analysis suggests that
  physical association between morphology and environment, may extend
  to much larger length scales than currently believed and
  the information theoretic framework presented here,
    can serve as a sensitive and useful probe of the assembly bias and
    large-scale environmental dependence of galaxy properties.
\end{abstract}


\section{Introduction}
The role of environment on galaxy formation and evolution is one of
the most complex issues in cosmology. The present day Universe is
filled with billions of galaxies which are distributed across a vast
network namely the `cosmic web' \citep{bond96} that stretches through
the Universe. This spectacular network of galaxies is made up of
interconnected filaments, walls and nodes which are encompassed by
vast empty regions. The galaxies broadly form and evolve in these four
types of environments inside the cosmic web. One can characterize the
environment of a galaxy with the local density at its location. The
role of local density on galaxy properties is well studied in
literature \citep{Oemler, dress, gotto, davis2, guzo,zevi,hog1, blan1,
  park1, einas2, kauffmann, mocine, koyama, bamford}. It is now well
known that the galaxy properties exhibit a strong dependence on the
local density of their environment. However the role of large-scale
environment on the formation and evolution of galaxies still remains a
debated issue.

The growth of primordial density perturbations leads to collapse of
dark matter halos in a hierarchical fashion. It is now widely accepted
following the seminal work by \citet{white78} that galaxies form at
the centre of the dark matter halos by radiative cooling and
condensation. One of the central postulates of the halo model
\citep{neyman, mo, ma, seljak, scocci2, cooray, berlind, yang1} is
that the halo mass determines all the properties of a
galaxy. But this need not be strictly true. The halos
are assembled through accretion and merger in different parts of the
cosmic web. Different accretion and merger histories of the halos
across different environments leads to assembly bias \citep{croton,
  gao07, musso, vakili} which manifests in the clustering of these
halos. The early-forming low mass halos in simulations
  are found to be more strongly clustered than the late-forming halos
  of similar mass.

  The presence of beyond halo mass effect in
  observations, is a matter of considerable debate due to conflicting
  results obtained by various studies on galactic conformity and
  assembly bias. A study \citep{zehavi11} of the colour and luminosity
  dependence of galaxy clustering in SDSS find that most observed
  trends can be explained by halo occupation distribution (HOD)
  modelling within a $\Lambda$CDM cosmology. \citet{alam19} study the
  dependence of clustering and quenching on the cosmic web using SDSS
  and show that the observed cosmic web dependence in the SDSS can be
  largely explained by HOD modelling without introducing any galaxy
  assembly bias. \citet{yan13} show that the galaxy properties do not
  depend on the tidal environment of the cosmic
  web. \citet{paranjape18} show that any observed dependence of galaxy
  properties on the tidal environment can be traced to those inherited
  from the assembly bias of their parent halos and additional effects
  of large-scale environment must be weak.  \citet{lin16} analyze the
  clustering of early and late forming halo samples using SDSS and
  find no significant evidence for assembly bias.  \citet{abbas07}
  show that environmental effects are also present in Poisson cluster
  models and the halo bias in these models are surprisingly similar to
  the standard models of halo bias \citep{mo}.  A number of
  observations suggest that the properties of satellite galaxies are
  strongly correlated with the central galaxy \citep{weinmann,
    kauffmann10, wang10, wangwhite}. \citet{tinker17} study the effect
  of halo formation history on quenching process in central galaxies
  and find a statistically significant impact at high masses and no
  impact at low masses. \citet{kauffmann13} find that the star
  formation rates in galaxies can be correlated upto 4
  Mpc. \citet{sin17} re-examine the nature of galactic conformity
  presented in \citet{kauffmann13} and find that such effects can
  arise due to selection biases. \citet{paranjape15} prescribed a
  tunable model within HOD framework to introduce varying levels of
  conformity in the mock galaxy catalogues and find no conclusive
  evidence of galaxy assembly bias on 4 Mpc. \citet{miyatake} study
  the halo bias of SDSS galaxy clusters using projected
  auto-correlation function and weak lensing and find that they differ
  by a factor of $1.5$, which could be a significant evidence of
  assembly bias.  \citet{zu17} study the possible origin of the
  discrepancy between the large scale halo bias of galaxy clusters
  \citep{miyatake} and find that these differences mostly arise due to
  projection effects. A recent work by \citet{kerscher} reported the
  existence of galactic conformity out to 40 Mpc. \citet{montedorta}
  analyze LRGs from SDSS-III BOSS survey and find a strong
  observational evidence of assembly bias.

Some other works \citep{lupa, scudder, pandey2, pandey3, darvish,
  filho} report significant dependence of the luminosity, star
formation rate and metallicity of galaxies on the large-scale
environment. A recent study by \citet{Lee} show that both the least
and most luminous elliptical galaxies in sheetlike structures inhabit
the regions with highest tidal coherence. It has been shown that the
large-scale environments in the cosmic web influence the mass, shape
and spin of dark matter halos \citep{hahn1, hahn2}. A number of
studies \citep{trujillo, erdogdu, paz, jones, tempel1,tempel2} suggest
alignment of halo shapes and spins with filaments which can extend
upto $40$ Mpc \citep{chen}.  In a recent study \citet{pandey17} use
information theoretic measures to show that the galaxy morphology and
environment in the SDSS exhibit a synergic interaction at least upto a
length scale of $ \sim 30 \hmpc$. A more recent study \citep{pandey20}
find that the fraction of red galaxies in sheets and filaments
increases with the size of these large-scale structures. Any such
large-scale correlations beyond the extent of the dark matter halo are
unlikely to be explained by direct interactions between them. All
these observations suggest that the role of environment on galaxy
formation and evolution may not be limited to local density alone. The
morphology and coherence of large-scale patterns in the cosmic web may
play a significant role in determining the galaxy properties and their
evolution.

\citet{pandey17} use mutual information to quantify the large-scale
environmental dependence of galaxy morphology. They find a non-zero
mutual information between morphology of galaxies and their
environment which decreases with increasing length scales but remains
non-zero throughout the entire length scales probed. In the present
work, we would like to test the statistical significance of mutual
information between morphology and environment and study its validity
and effectiveness as a measure of large-scale environmental dependence
of galaxy properties for future studies.

We propose a method where we destroy the correlation between
morphology and environment by randomizing the morphological
classification and measure the mutual information to test its
statistical significance. We also divide the data into cubes and
shuffle them around many times to test how the mutual information
between morphology and environment are affected by the shuffling
procedure. We carry out these tests using data from the Galaxy Zoo
database \citep{lintott08}. Further, we carry out a controlled test
using a semi-analytic galaxy catalogue \citep{henriques15} based on
the Millennium simulation \citep{springel05}. The galaxies in these
mock datasets are selectively assigned morphology based on their local
density. We measure the mutual information between morphology and
environment in each case and try to understand the statistical
significance of mutual information in the present context. The goal of
the present analysis is to explore the potential of mutual information
as a statistical measure to reveal the large-scale correlations
between environment and morphology if any.

A $\Lambda$CDM cosmological model with $\Omega_{m0}=0.315$,
$\Omega_{\Lambda0}=0.685$ and $h=0.674$ \citep{planck18} is used to
convert redshifts to distances throughout the analysis.

\section{DATA}

\subsection{SDSS DR16}

We use data from the $16^{th}$ data release \cite{ahumada19} of Sloan
Digital Sky Survey (SDSS) \cite{york00}. DR16 is the final data
release of the fourth phase of SDSS which covers more than nine
thousand square degrees of the sky and provides spectral information
for more than two million galaxies. This includes an accumulation of
data collected for new targets as well as targets from all prior data
releases of SDSS. The data is downloaded through {\it SciServer:
  CASjobs}\footnote{https://skyserver.sdss.org/casjobs/} which is a
SQL based interface for public access. We identify a contiguous region
within $0^{\circ} \leq \delta \leq 60^{\circ}$ \& $ 135^{\circ} \leq
\alpha \leq 225^{\circ}$ and select all galaxies with the apparent
r-band Petrosian magnitude limit $m_r<17.77$ within that region. Here
$\alpha$ and $\delta$ are the right ascension and declination
respectively. We combine the three tables {\it SpecObjAll}, {\it
  Photoz} and {\it ZooSpec} of SDSS database to get the required
information about each of these selected galaxies. We retrieve the
spectroscopic and photometric information of galaxies from the {\it
  SpecObjAll} and {\it Photoz} tables respectively. The {\it ZooSpec}
table provides The morphological classifications for the SDSS galaxies
from the Galaxy Zoo project\footnote{http://zoo1.galaxyzoo.org}.
Galaxy zoo \citep{lintott08,lintott11} is a platform where millions of
registered volunteers vote for visual morphological classification of
galaxies. These votes contribute in identification of galaxy
morphologies through a structured algorithm. The galaxies in galaxy
zoo are flagged as {\it spiral}, {\it elliptical} or {\it uncertain}
depending on the vote fractions. We only consider the galaxies which
are flagged as {\it spiral} or {\it elliptical} with debiased vote
fraction $>0.8$ \citep{bamford}. These cuts yield a total $136155$
galaxies within redshift $z<0.3$. We then construct a volume limited
sample using a r-band absolute magnitude cut $M_r \leq -20.5$.  This
provides us $44049$ galaxies within $z<0.096$. The present analysis
requires a cubic region. We extract a cubic region of side $145 \hmpc$
from the volume limited sample which contains $14558$ galaxies. The
resulting datacube consists of $11171$ spiral galaxies and $3387$
elliptical galaxies.  The mean intergalactic separation of the
galaxies in this sample is $\sim 6 \hmpc$.

\begin{figure*}
\resizebox{7.6 cm}{!}{\rotatebox{0}{\includegraphics{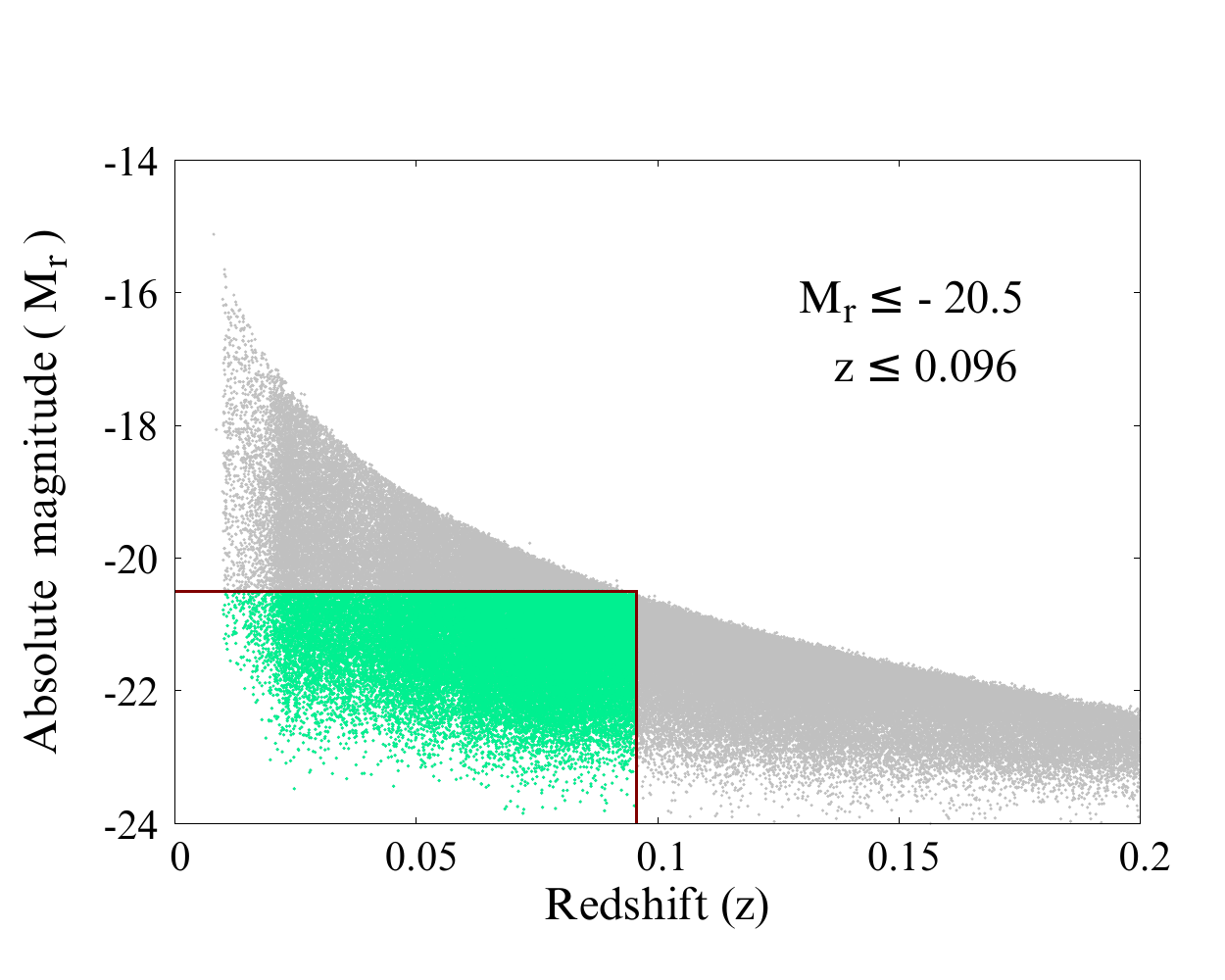}}}    \hspace{0.5cm}
\resizebox{7.6 cm}{!}{\rotatebox{0}{\includegraphics{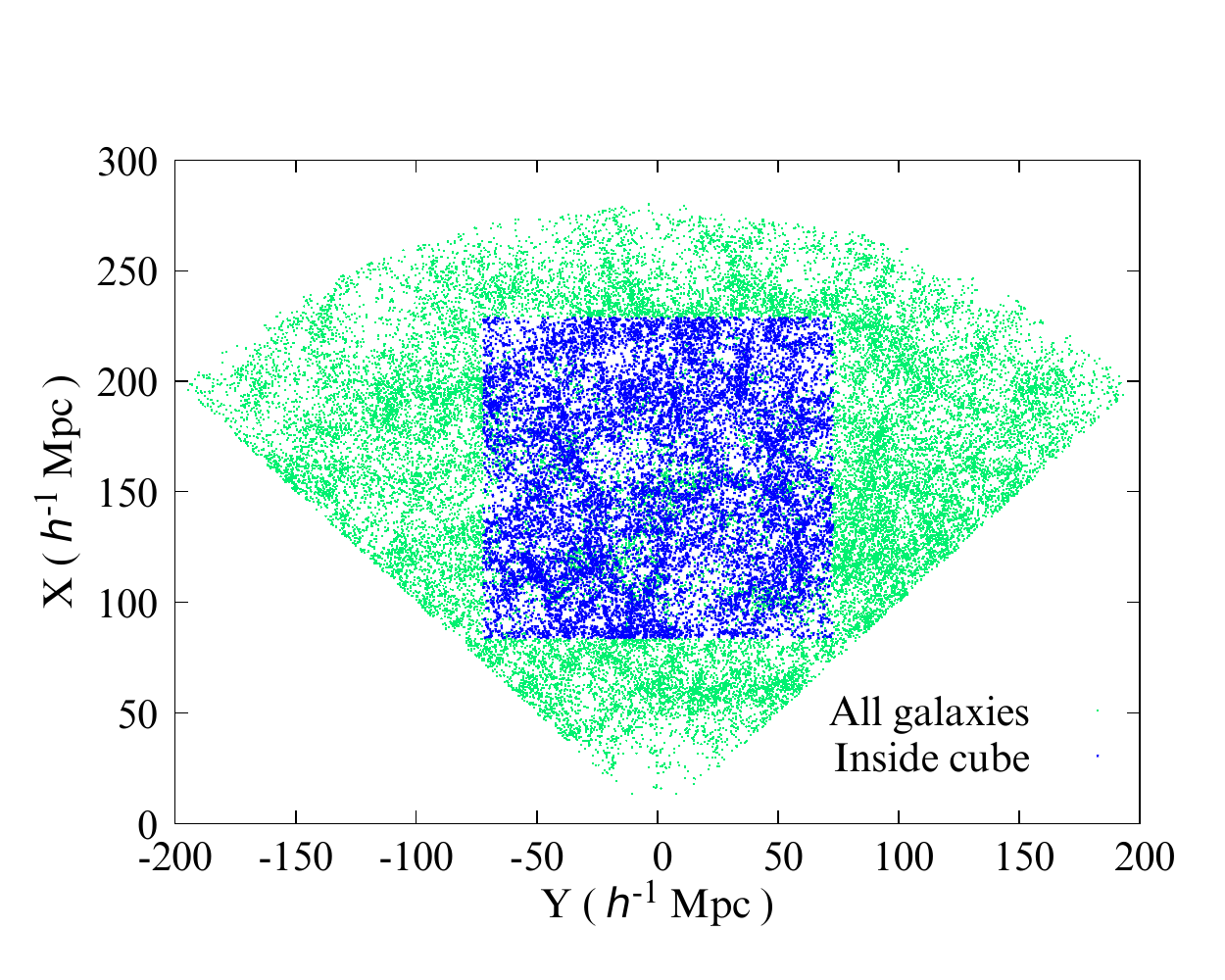}}}  \\
\resizebox{7.4 cm}{!}{\rotatebox{0}{\includegraphics{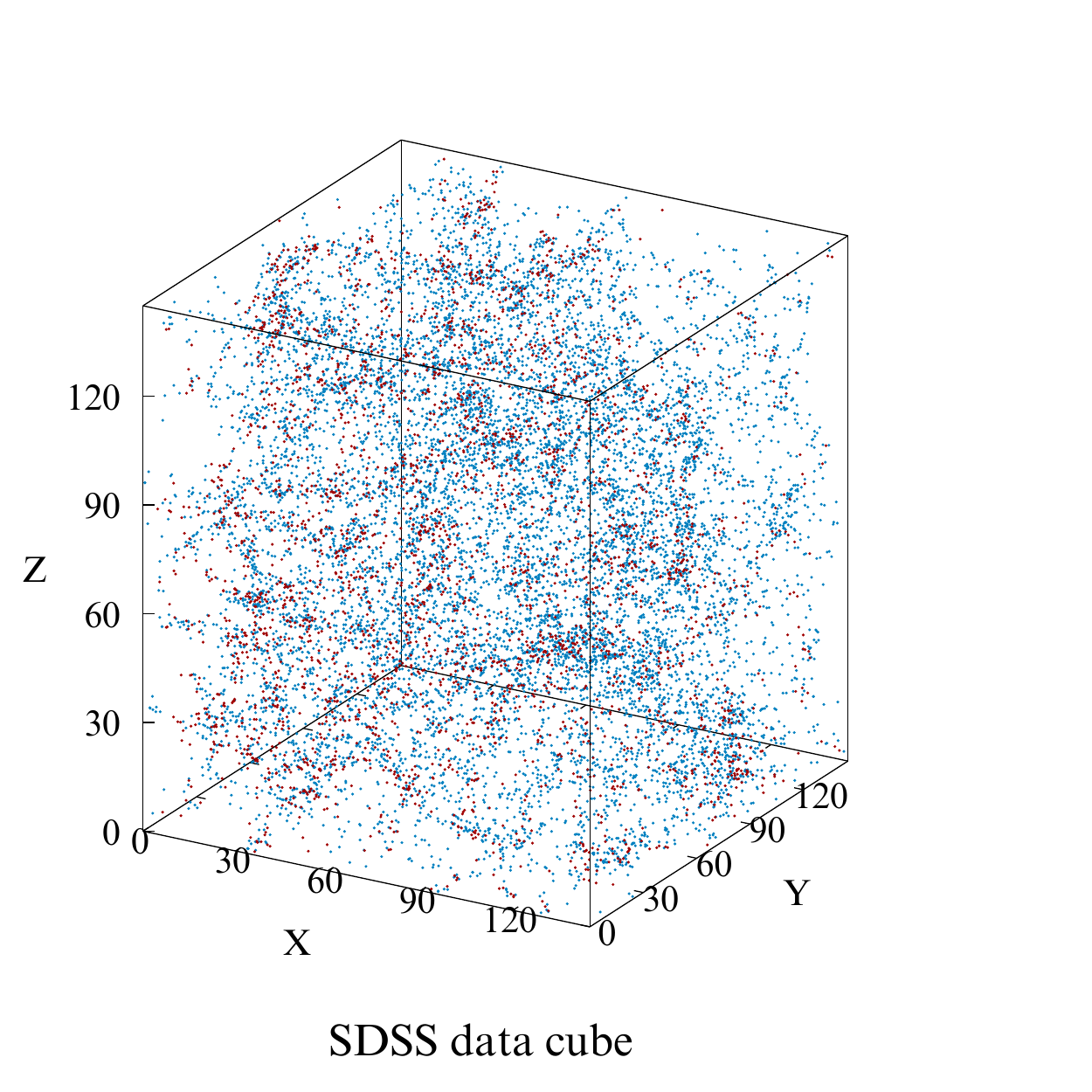}}} \hspace{0.5cm}
\resizebox{7.0 cm}{!}{\rotatebox{0}{\includegraphics{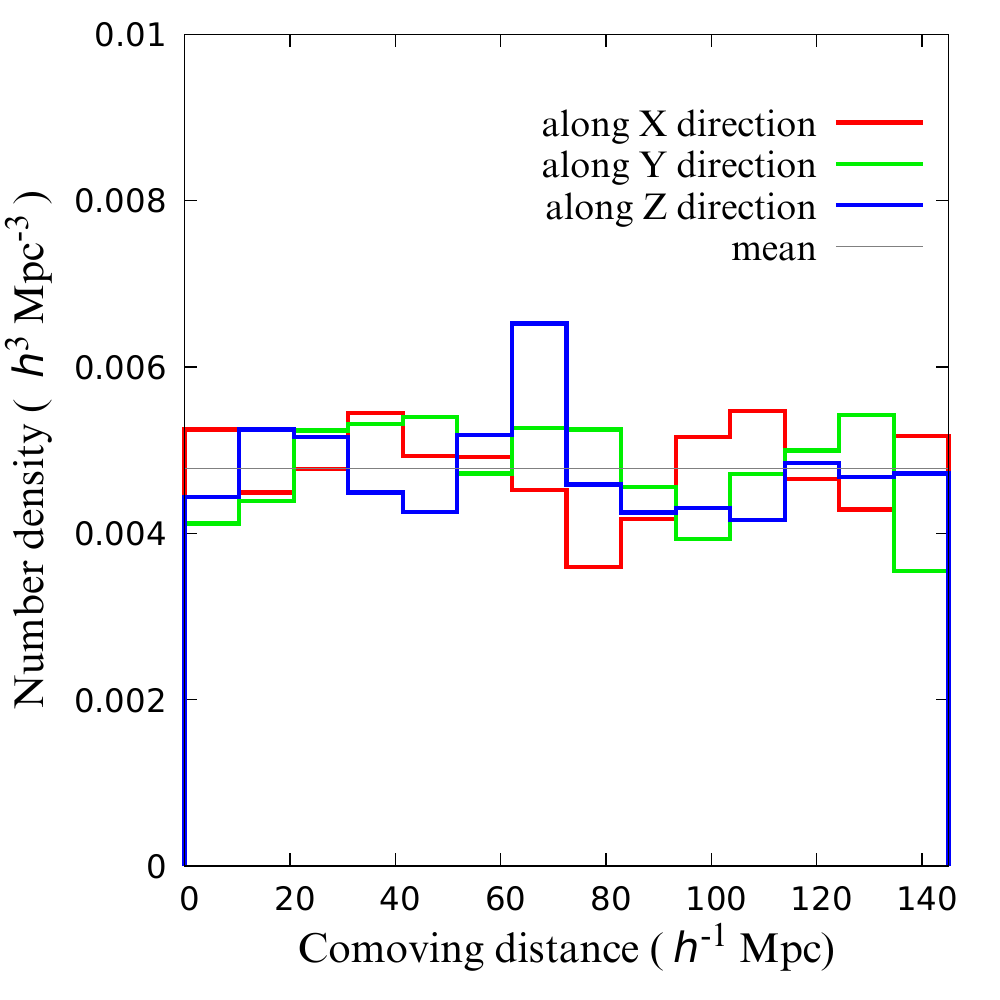}}} \\
\caption{The top left panel shows the definition of the volume limited
  sample in the redshift-absolute magnitude plane. The top right panel
  shows the projected view of the galaxies in the entire volume
  limited sample (green dots) and those inside the cubic region (blue
  dots). The bottom left panel shows the distributions of spirals
  (blue dots) and ellipticals (brown dots) in the extracted datacube
  from the volume limited sample. The bottom right panel shows the
  variation in number density inside the datacube along each of the
  3-axes. The number densities are computed in slices of thickness
  $10.36 \hmpc$.}
  \label{fig:sample}
\end{figure*}

\begin{figure*} 
  \resizebox{7.6 cm}{!}{\rotatebox{0}{\includegraphics{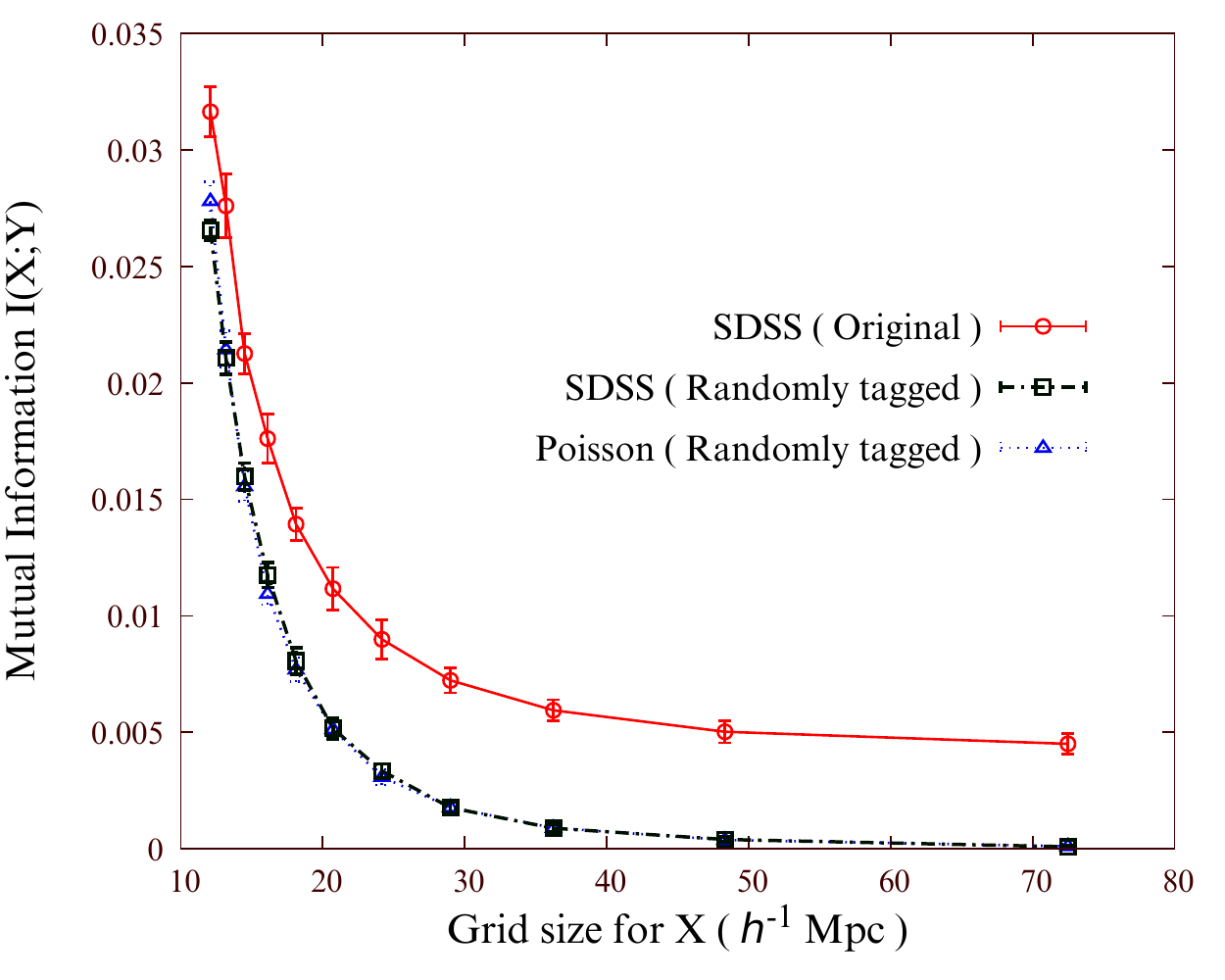}}}
  \resizebox{7.6 cm}{!}{\rotatebox{0}{\includegraphics{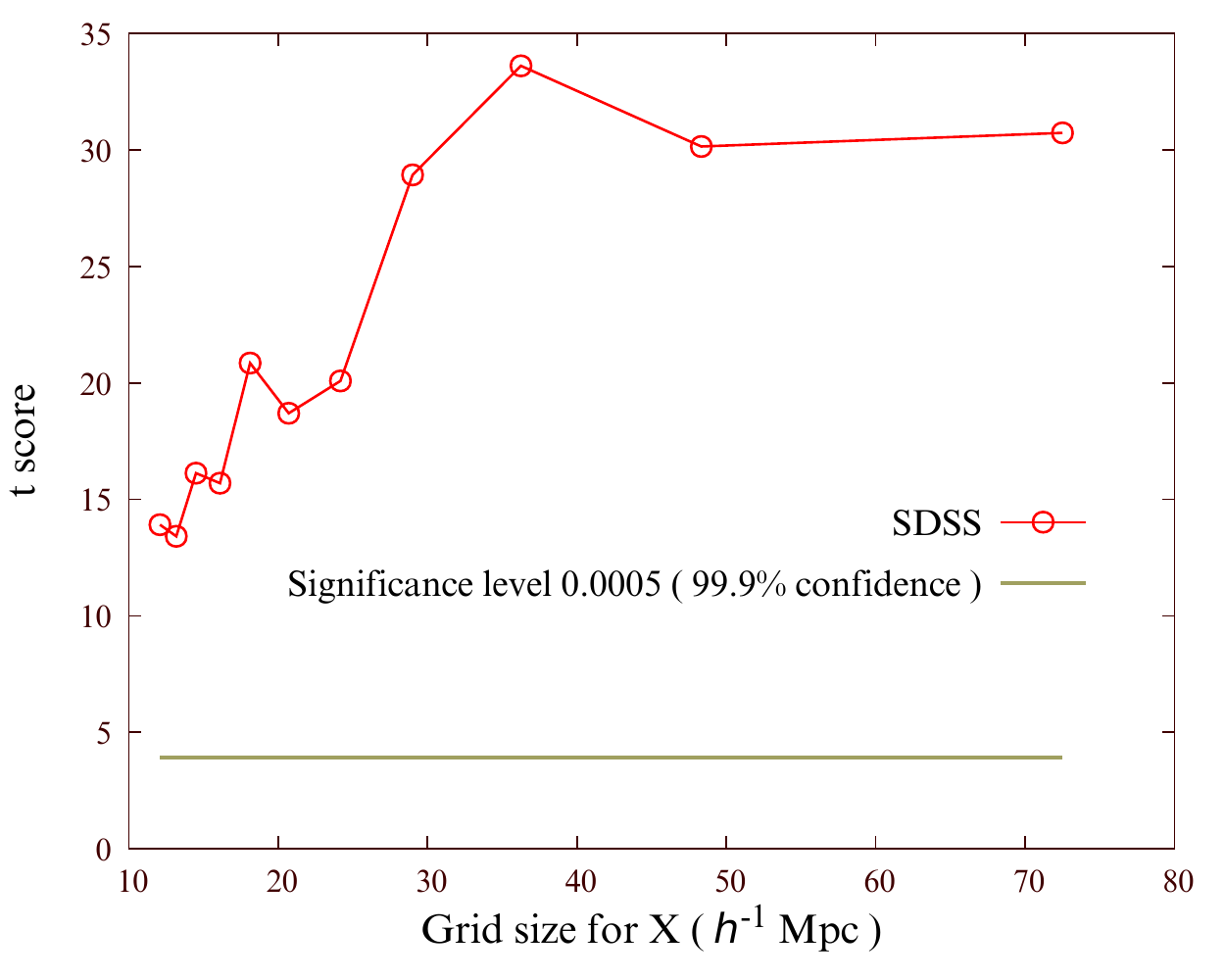}}}
\caption{The left panel of this figure shows the mutual information
  $I(X;Y)$ as a function of length scales for the original SDSS
  datacube and the SDSS datacube where the morphological information
  of galaxies are randomized. The results for mock Poisson
  distribution with randomly assigned morphology are also shown
  together for a comparison. The $1-\sigma$ errorbars for the original
  SDSS data are estimated using $10$ jack-knife samples drawn from the
  same dataset. For the SDSS random and Poisson random datasets each,
  we estimate the $1-\sigma$ errobars using 10 different
  realizations. The right panel of this figure shows the t score as a
  function of length scales, obtained from a $t$-test which compares
  the SDSS galaxy distribution with randomized morphological
  classification to the SDSS galaxy distribution with actual
  morphological classification.}
  \label{fig:Imxy_ran}
\end{figure*}

\begin{figure*} 
\resizebox{8 cm}{!}{\rotatebox{0}{\includegraphics{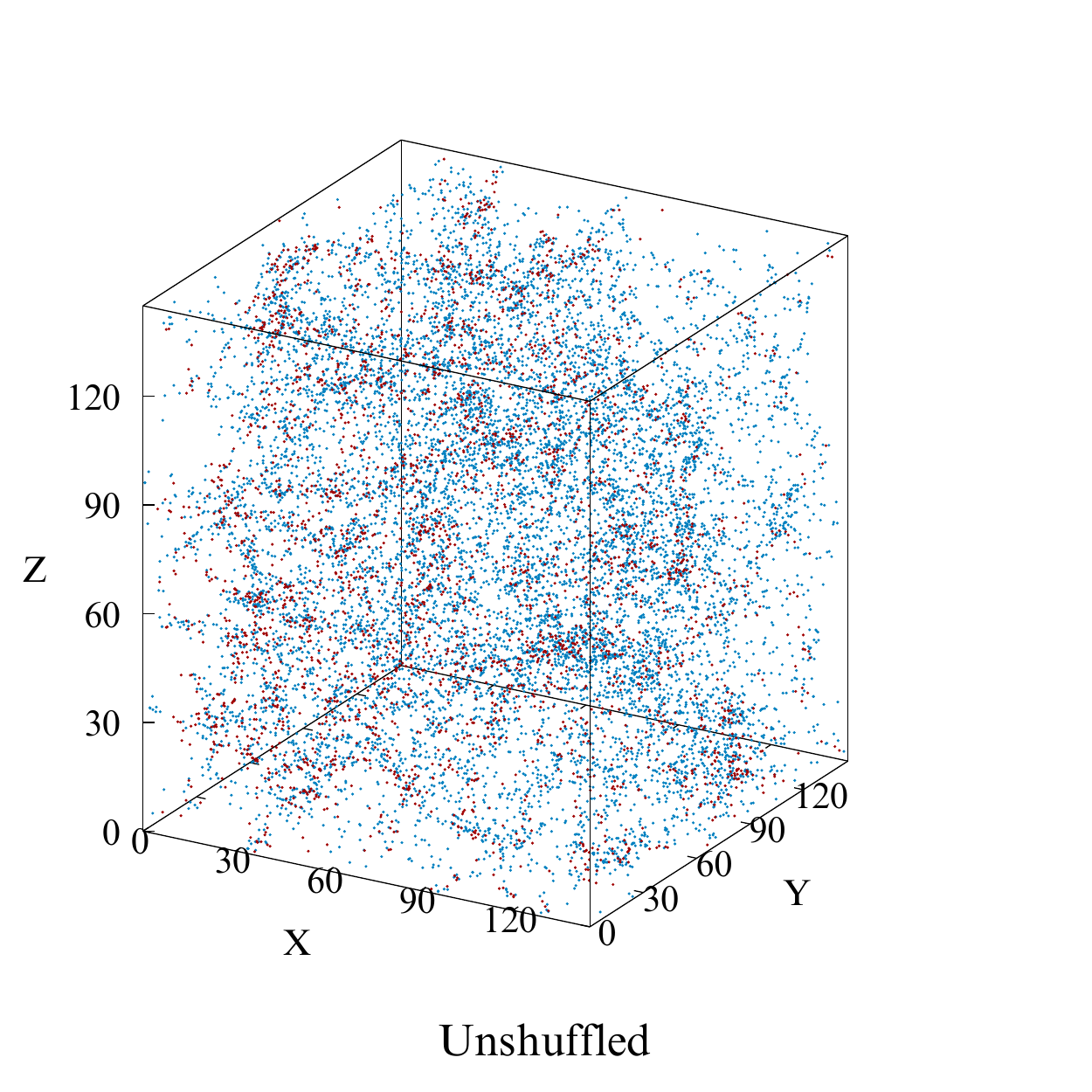}}}
\resizebox{8
  cm}{!}{\rotatebox{0}{\includegraphics{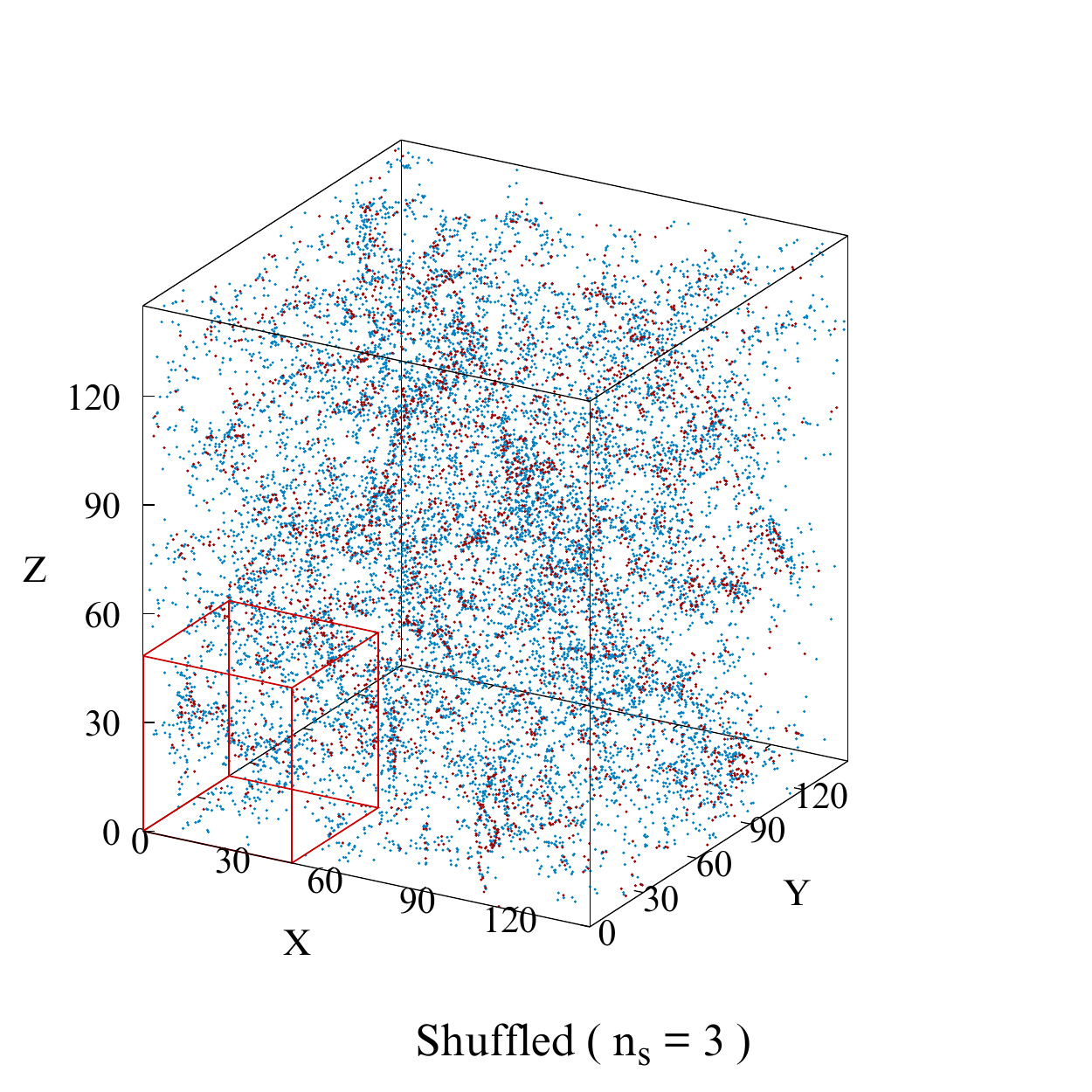}}}
\\ \resizebox{8
  cm}{!}{\rotatebox{0}{\includegraphics{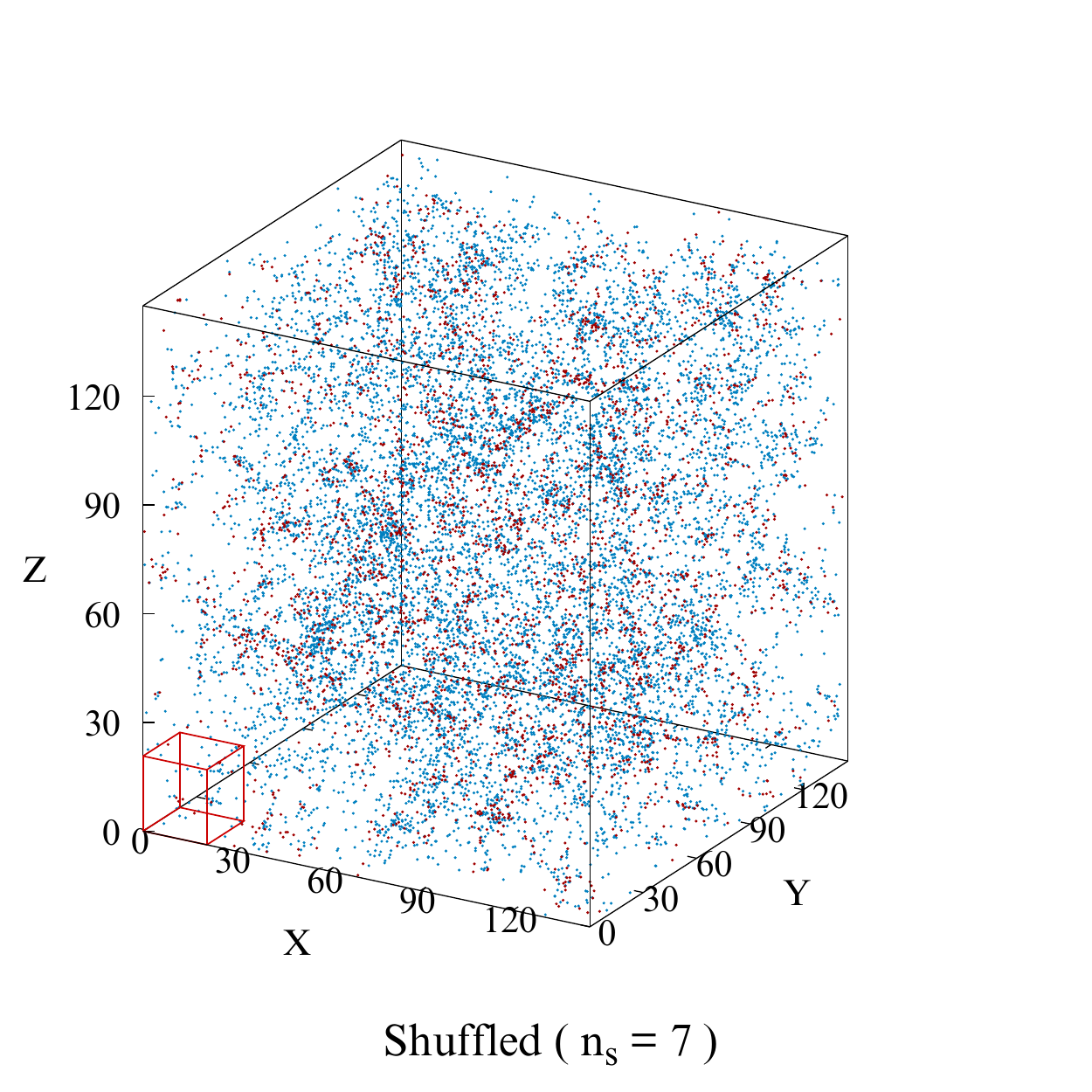}}}
\resizebox{8
  cm}{!}{\rotatebox{0}{\includegraphics{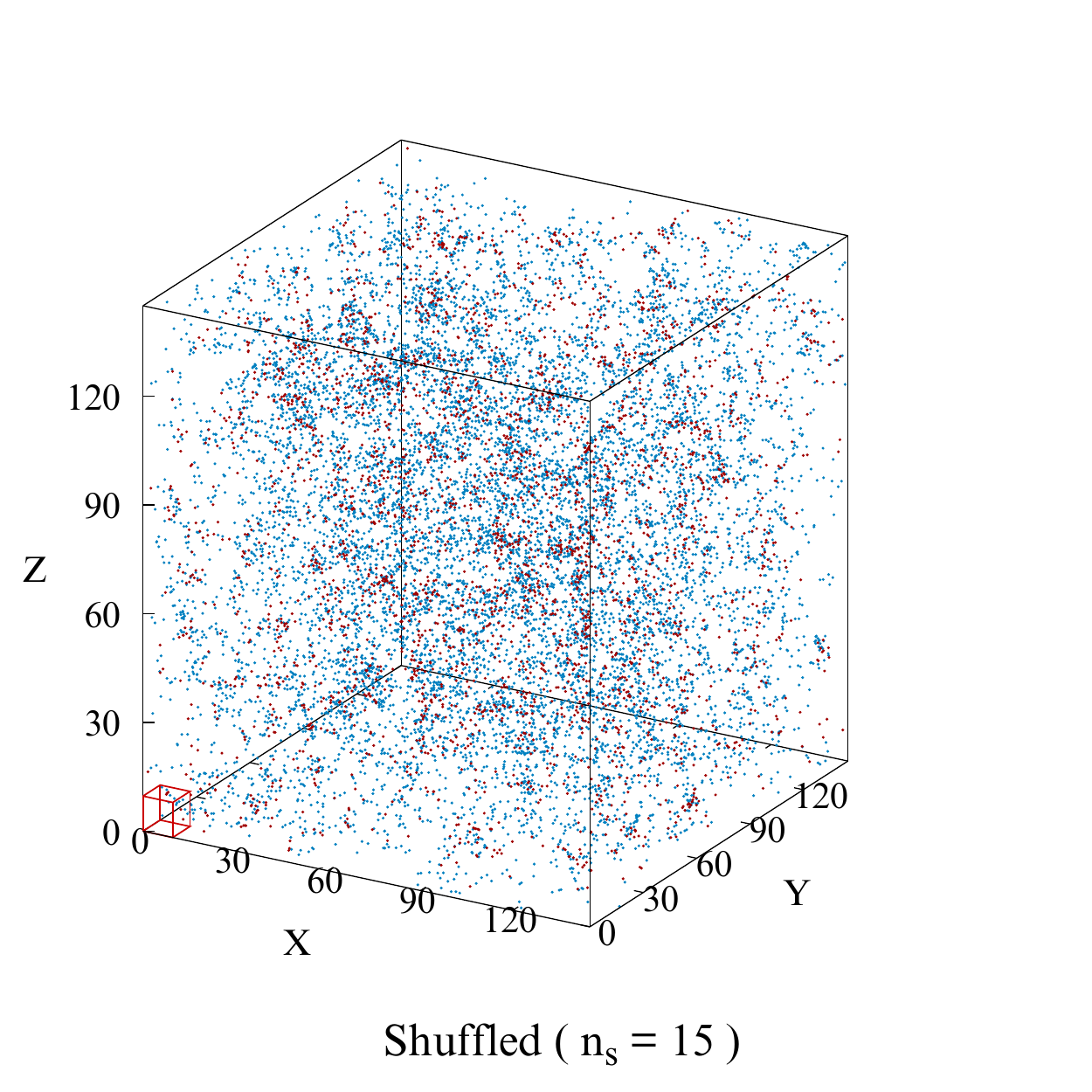}}}
\caption{This figure shows the distributions of spirals (blue dots)
  and ellipticals (brown dots) in the original unshuffled SDSS
  datacube along with one realization of shuffled datacube for three
  different values of shuffling lengths ($l_s$). The value of $l_s$ is
  decided by $n_s$ which is the number of subcubes that would fit
  along each dimension. The size of shuffling units in each case is
  shown with a subcube (in red) at a corner of the respective shuffled
  realization.}
  \label{fig:3Dv_shuf}
\end{figure*}

\begin{figure*} 
  \resizebox{7.6 cm}{!}{\rotatebox{0}{\includegraphics{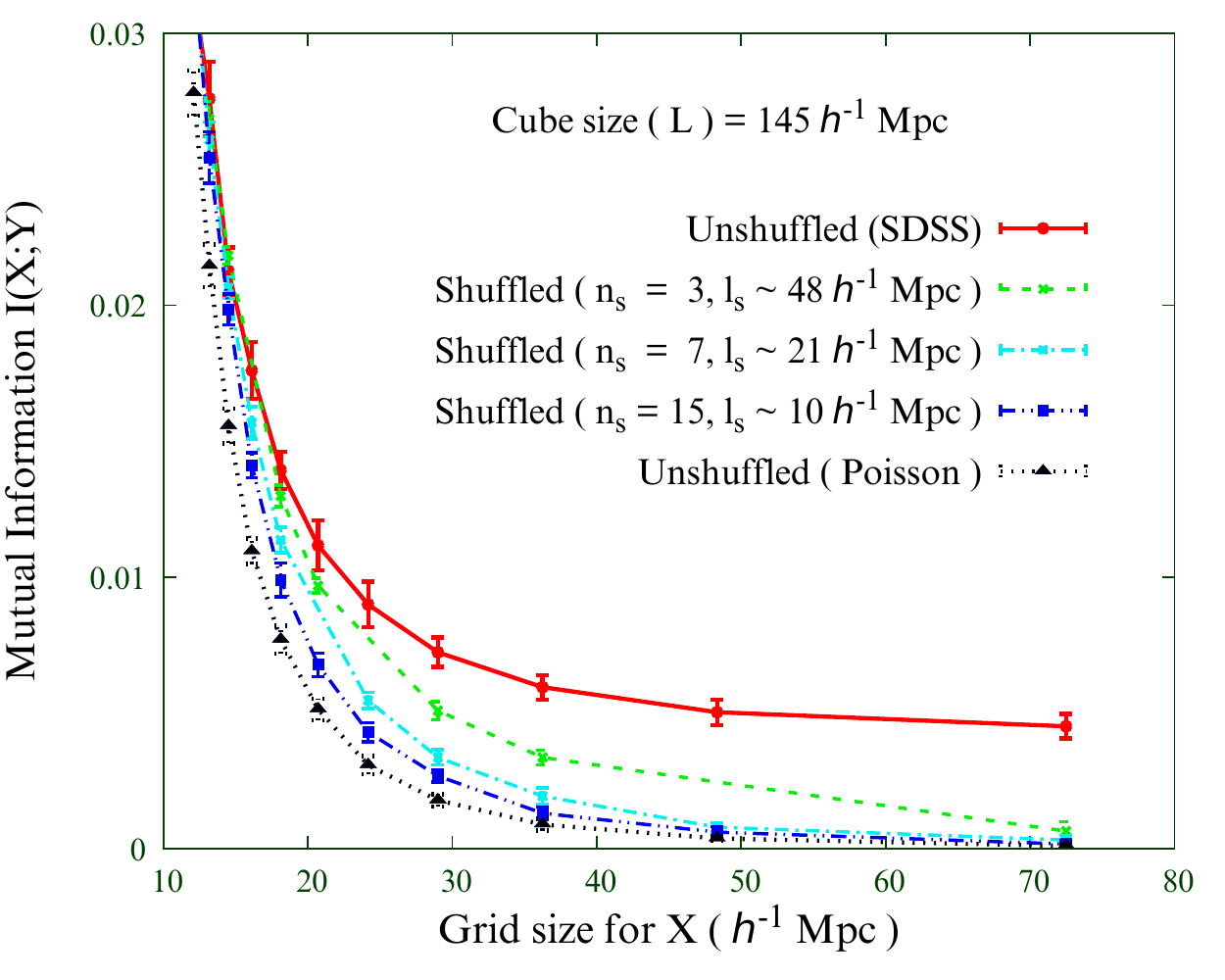}}} 
  \resizebox{7.6 cm}{!}{\rotatebox{0}{\includegraphics{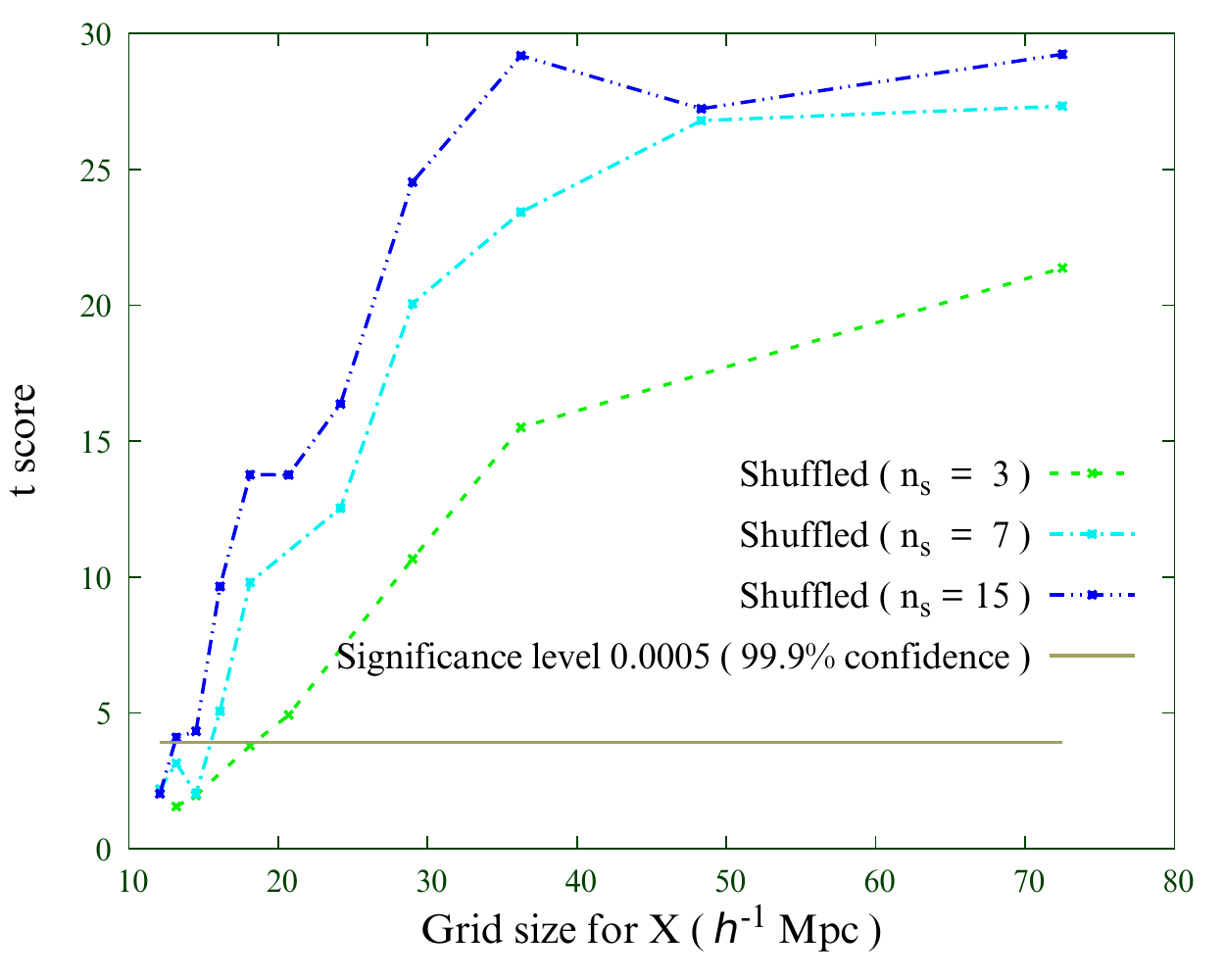}}}
\caption{The left panel of this figure shows the mutual information
  $I(X;Y)$ as a function of length scales in the unshuffled SDSS
  datacube along with that from the shuffled realizations with three
  different shuffling length. The $1-\sigma$ errorbars shown for the
  unshuffled SDSS data are obtained from $10$ jack-knife samples drawn
  from the same dataset. For the SDSS shuffled datasets and Poisson
  random datasets each, the $1-\sigma$ errobars are estimated using 10
  different realizations. For each shuffling length, the grid sizes
  are chosen so that they are not equal or integral multiples of the
  shuffling length and the vice versa. The right panel of this figure
  shows the t score at different length scales, obtained from a $t$
  test comparing the shuffled distributions with the original
  unshuffled galaxy distribution from SDSS.}
  \label{fig:Imxy_shuf}
\end{figure*}

\begin{figure*} 
\resizebox{8 cm}{!}{\rotatebox{0}{\includegraphics{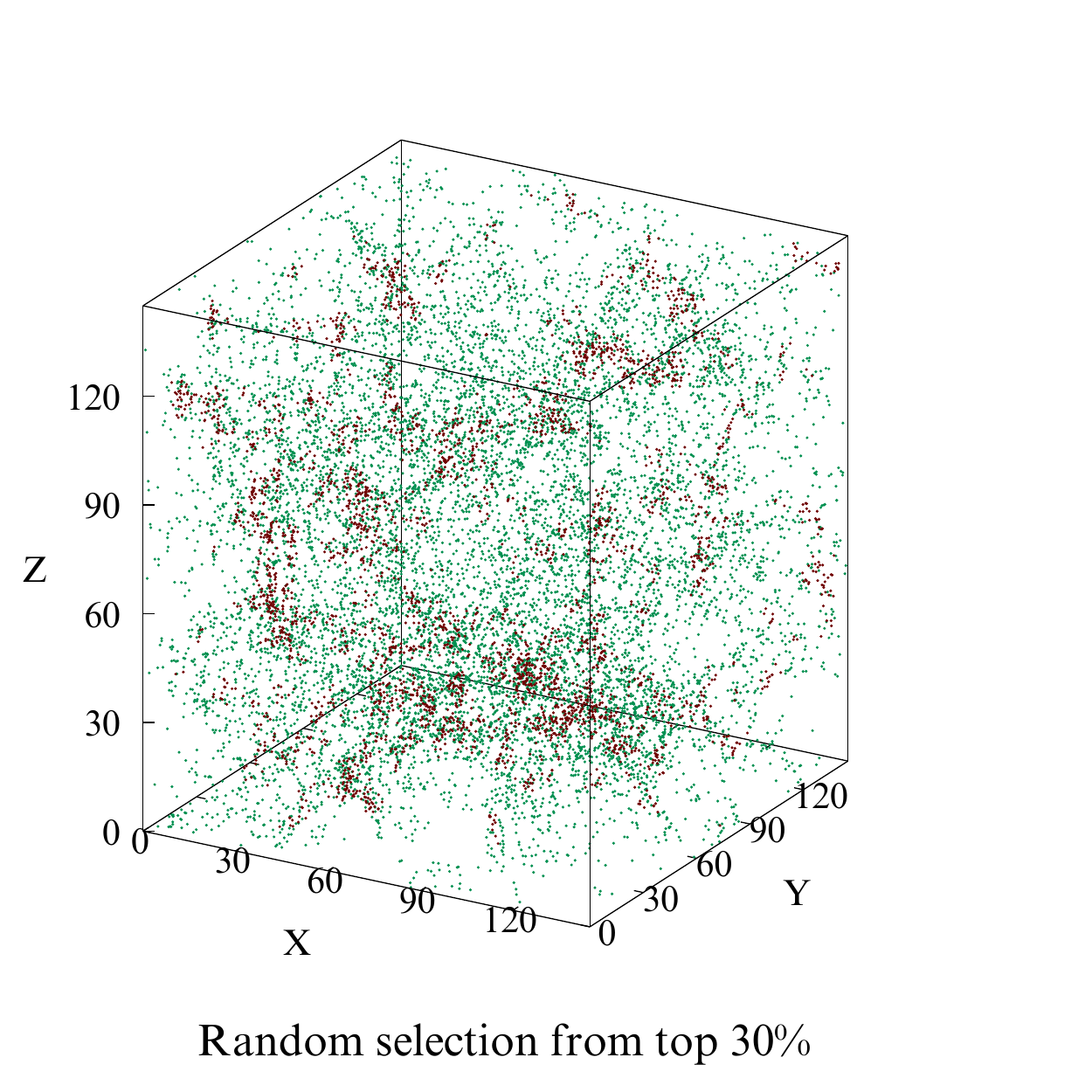}}} 
\resizebox{8 cm}{!}{\rotatebox{0}{\includegraphics{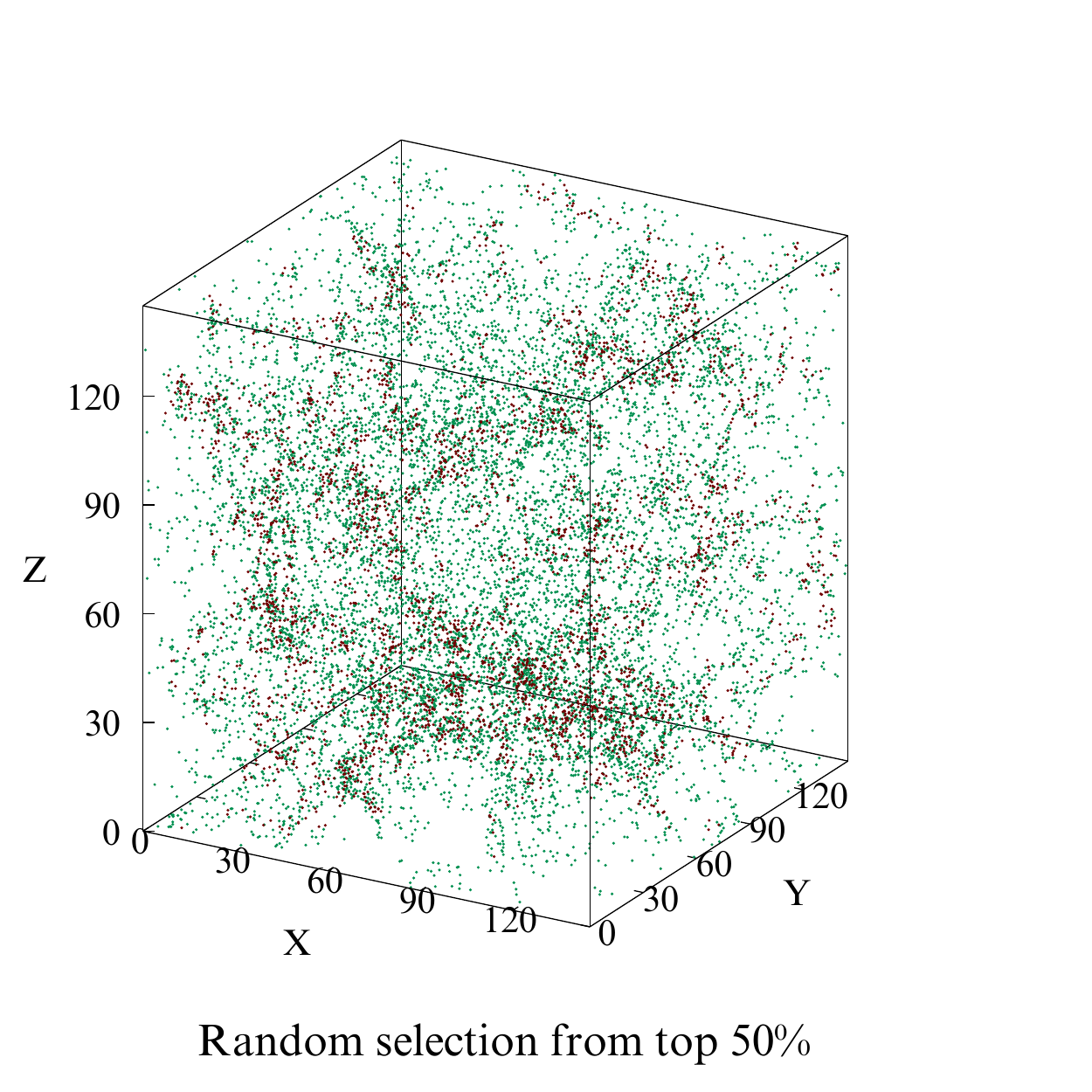}}} 
\resizebox{8 cm}{!}{\rotatebox{0}{\includegraphics{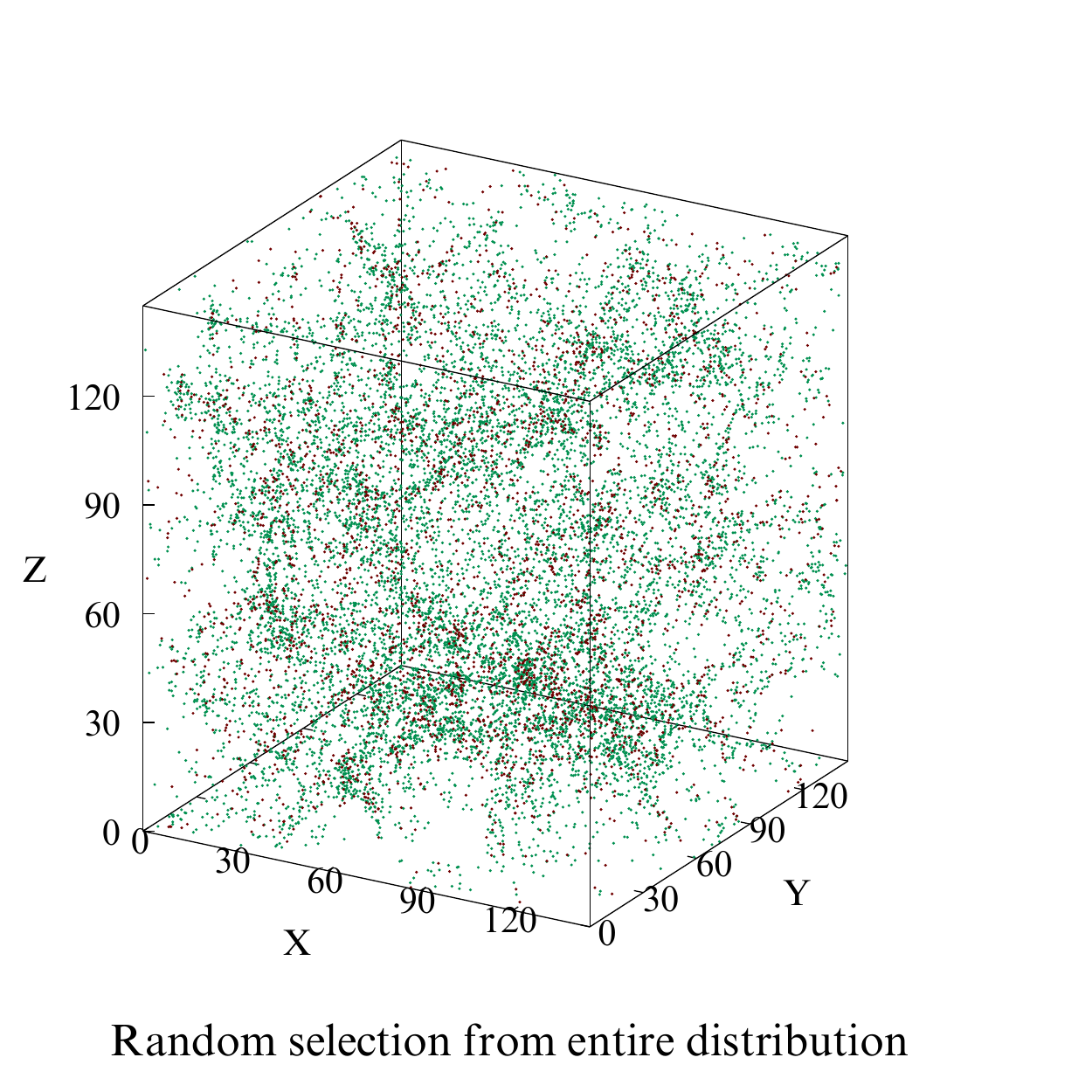}}}
\caption{This figure shows the distributions of spirals (greenish dot)
  and ellipticals (brown dot) in a realization of the mock SDSS
  datacube from SAM where the galaxies are assigned morphology based
  on the density at their locations. The three datacubes corresponds
  to three different schemes for density dependent morphology
  assignment.}
  \label{fig:3Dv_mill}
\end{figure*}

\begin{figure*} 
  \resizebox{7.6 cm}{!}{\rotatebox{0}{\includegraphics{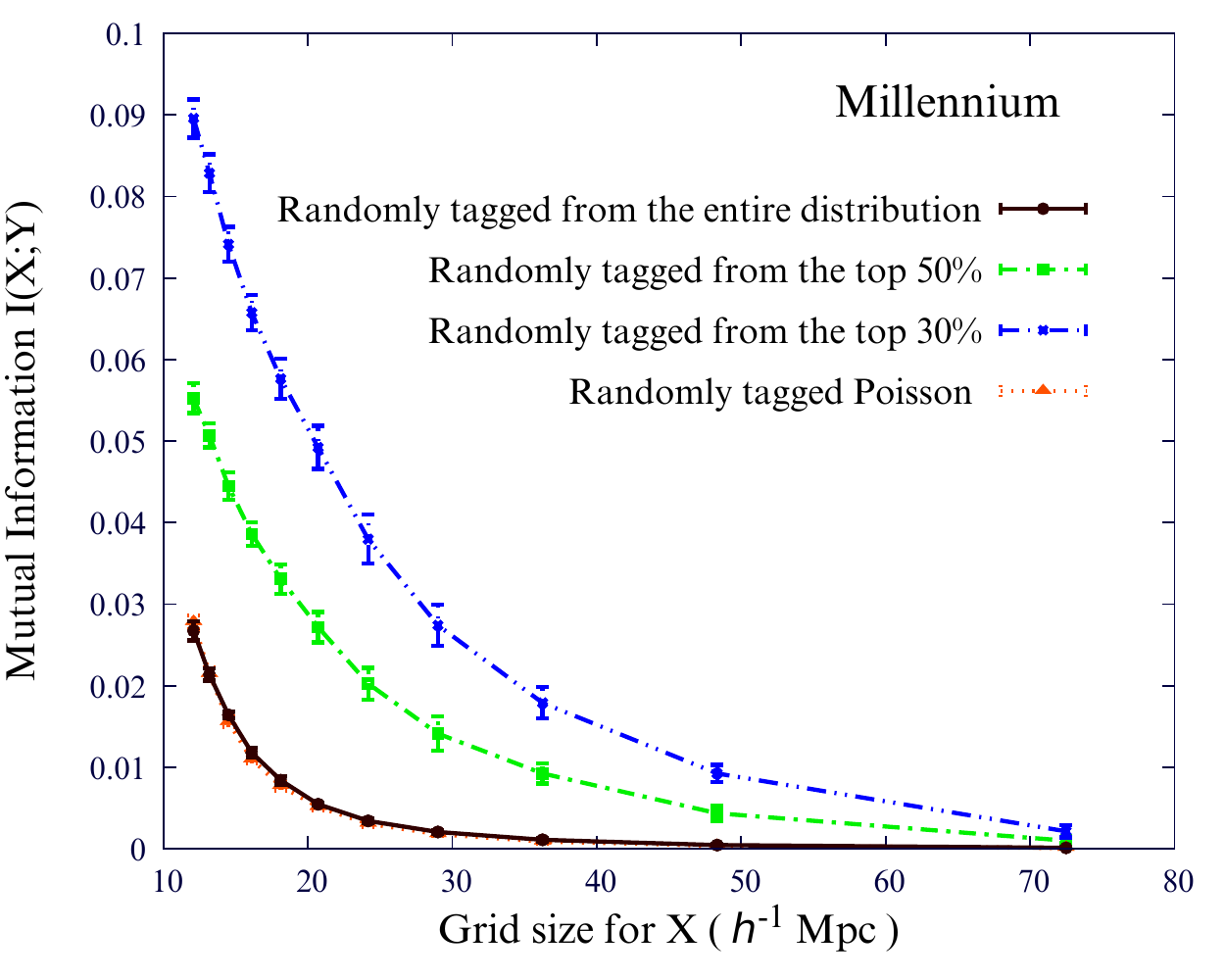}}} 
  \resizebox{7.6 cm}{!}{\rotatebox{0}{\includegraphics{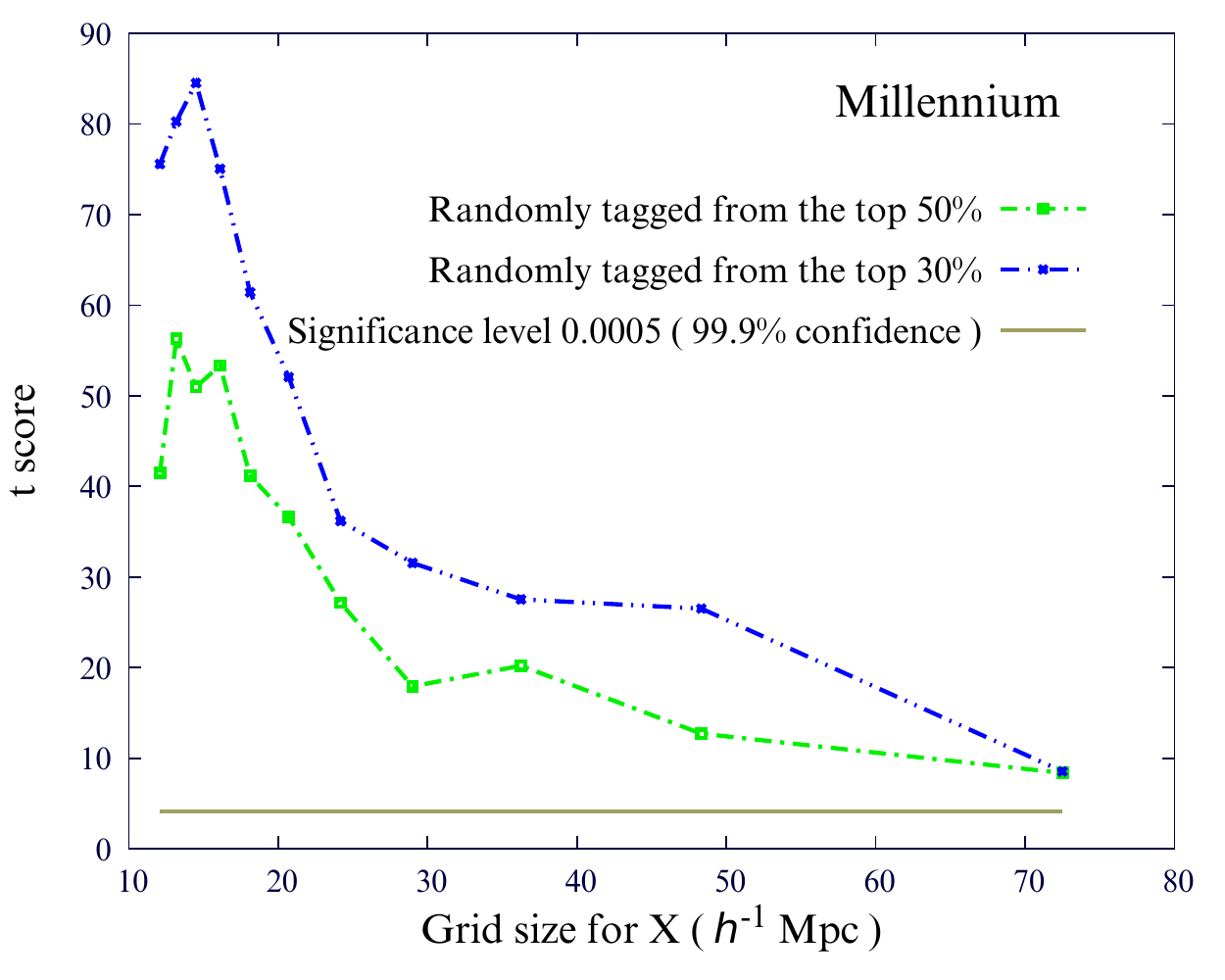}}}
\caption{The left panel of this figure shows mutual information
  $I(X;Y)$ as a function of length scales for different
  morphology-density relations. $1-\sigma$ errorbars for the
  Millennium galaxies are estimated using data from 8 non-overlapping
  mock datacubes from the SAM catalogue. The $1-\sigma$ errorbars
  corresponding to the Poisson dataset are estimated using 8 mock
  datacubes containing random distributions. The right panel of this
  figure shows the t score at different length scales, obtained from
  $t$ test where we compared the distributions with density dependent
  morphological tagging to that without any density dependence of
  morphology.}
  \label{fig:Imxy_mill}
\end{figure*}

\subsection{Millennium Run Simulation}

 Galaxy formation and evolution involve many complex physical
 processes such as gas cooling, star formation, supernovae feedback,
 metal enrichment, merging and morphological evolution. The semi
 analytic models (SAM) of galaxy formation
 \citep{white91,kauff1,cole1,bagh,somervil,benson} is a powerful tool
 which parametrise these complex physical processes in terms of simple
 models following the dark matter merger trees over time and finally
 provide the statistical predictions of galaxy properties at any given
 epoch. In the present work, we use the data from a semi analytic
 galaxy catalogue \citep{henriques15} derived from the Millennium run
 simulation (MRS) \citep{springel05}. \citet{henriques15} updated the
 Munich model of galaxy formation using the values of cosmological
 parameters from PLANCK first year data. This model provides a
 better fit to the observed stellar mass functions and reproduce the
 recent data on the abundance and passive fractions of galaxies over
 the redshift range $0 \le z \le 3$ better than the other models. We
 use SQL to extract the required data from the Millennium
 database \footnote{https://www.mpa.mpa-garching.mpg.de/millennium/}.
 We use the peculiar velocities of the Millennium galaxies to map them
 in redshift space and extract all the galaxies with $M_r \leq -20.5
 $. Finally we construct 8 mock SDSS datacubes of side $145 \hmpc$
 each containing a total $14558$ galaxies.

\subsection{Random distributions}

We simulate 10 Poisson distributions each within a cube of side $145
\hmpc$. $14558$ random data points are generated within each of the 10
datacubes. For each cube, we randomly label $3387$ points as
elliptical and rest of the points are labelled as spirals. The number
of galaxies and the ratio of spirals to ellipticals in these random
data sets are identical to that observed in the original SDSS
datacube.

\section{Method of analysis}

\subsection{Mutual information between environment and morphology}
We consider a cubic region of side $L \hmpc$ extracted from the volume
limited sample prepared from SDSS DR16. We subdivide the entire cube
into $N_{d}$ number of $d \hmpc \times d \hmpc \times d \hmpc$
voxels. We define a discrete random variable $X$ with $N_d$ outcomes
$\{ X_i:i=1,...N_d \}$. The probability of finding a randomly selected
galaxy in the $i^{th}$ voxel is $p(X_i)=\frac{N_i}{N}$, where $N_i$ is
the number of galaxies in the $i^{th}$ voxel and $N$ is the total
number of galaxies in the cube. The random variable $X$ thus defines
the environment of a galaxy at a specific length scale $d \hmpc$.

The information entropy \citep{shannon48} associated with the random
variable $X$ at scale $d$ is given by
\begin{eqnarray}
H(X)& = &-\sum_{i=1}^{N_d} p(X_i) \log p(X_i) \nonumber \\ &=&\log N -
\frac{\sum_{i=1}^{N_d} N_i \log N_i}{N}
  \label{eqn:Hx}
\end{eqnarray}

We use another variable $Y$ to describe the morphology of the
galaxies. We have only considered the galaxies with a classified
morphology and hence there are only two possible outcomes:
{\it{spiral}} or {\it{elliptical}}. If the cube consists of $N_{sp}$
spiral galaxies and $N_{el}$ elliptical galaxies then the information
entropy associated with $Y$ will be
\begin{eqnarray}
H(Y)& = &- \left( \frac{N_{sp}}{N} \log \frac{N_{sp}}{N} + \frac{N_{el}}{N} \log \frac{N_{el}}{N} \right) \nonumber \\
&=& \log N- \frac{ N_{sp} \log N_{sp} + N_{el} \log N_{el}}{N}
  \label{eqn:Hy}
\end{eqnarray}

Now having the prior information about the morphology of each of the
galaxies one can determine the mutual information between morphology
of the galaxies and their environment.

The mutual information $I(X;Y)$ between environment and morphology is ,
\begin{eqnarray}
I(X;Y) & = & \sum^{N_{d}}_{i=1} \sum^{2}_{j=1} \, p(X_i,Y_j) \, \log\, \frac{p(X_i,Y_j)}{p(X_i)p(Y_j)} \\
\nonumber & = & H(X)+H(Y)-H(X,Y)
\label{eq:Ixy}
\end{eqnarray}

$H(X)$ and $H(Y)$ are the individual entropy associated with the
random variables $X$ and $Y$ respectively. The joint entropy
$H(X,Y)\leq H(X)+H(Y)$ where the equality holds only when $X$ and $Y$
are independent. The joint entropy is symmetric i.e. $H(X,Y)=H(Y,X)$.

If $N_{ij}$ is the number of galaxies in the $i^{th}$ voxel that
belongs to the $j^{th}$ morphological class ($j=1$ for spiral and
$j=2$ for elliptical), then the joint entropy $H(X,Y)$ is given by,
\begin{eqnarray}
H(X,Y) &=& -\sum_{i=1}^{N_d} \sum_{j=1}^{2} p(X_i,Y_j) \log p(X_i,Y_j) \nonumber \\
&=&\log N - \frac{1}{N}\sum_{i=1}^{N_d} \sum_{j=1}^{2} N_{ij} \log N_{ij} 
\label{eqn:Hxy}
\end{eqnarray}

where
\begin{eqnarray}
\sum_{i=1}^{N_d} \sum_{j=1}^{2} N_{ij}=N
\label{eqn:Nij}
\end{eqnarray}

Here $p(X_i,Y_j)=p(X_i|Y_j)p(Y_j)=\frac{N_{ij}}{N}$ is the joint
probability derived from the conditional probability using Bayes'
theorem.

The mutual information between two random variables measures the
reduction in uncertainty in the knowledge of one random variable given
the knowledge of other. A higher value of mutual information between
two random variables convey a greater degree of association between
the two random variables. One specific advantage of mutual information
over the traditional tools like covariance analysis is that it does
not require any assumptions regarding the nature of the random
variables and their relationship.

\subsection{Randomizing the morphological classification of galaxies}

We consider each of the SDSS galaxies in the datacube and randomly
identify them as spirals and ellipticals leaving aside their actual
morphology. We randomly pick $3387$ SDSS galaxies and tag them as
ellipticals. Rest of the galaxies in the SDSS datacube are labelled as
spirals. The number of spirals and ellipticals in the resulting
distribution thus remains same as the original distribution.

We generate 10 such datacubes with randomly assigned galaxy morphology
from the original SDSS datacube and measure the mutual information
between environment and morphology in each of them. We would like to
compare the mutual information $I(X;Y)$ measured in the original SDSS
data with that from the SDSS dataset with randomly assigned morphology
to study the statistical significance of $I(X;Y)$ and its scale
dependence.

\subsection{Shuffling the spatial distribution of galaxies}

We divide the SDSS datacube of side $L \hmpc$ into $N_c=n_s^3$ smaller
subcubes of size $l_s=\frac{L}{n_s} \hmpc$. Each of these smaller
subcubes along with all the galaxies within them are rotated around
three different axes by different angles which are random multiples of
$90^{\circ}$. The rotated subcubes are then randomly interchanged with
any other subcubes inside the datacube. This process of arbitrary
rotation followed by random swapping is repeated for $100 \times N_c$
times to generate a {\it{Shuffled}} realization \citep{bhav} from the
original SDSS datacube. We carry out the shuffling procedure for three
different choices $n_s=3$, $n_s=7$ and $n_s=15$ which corresponds to
shuffling length $l_s=48.33 \hmpc$, $l_s= 20.71\hmpc$ and $l_s=
9.67\hmpc$ respectively. We generate 10 shuffled realizations for each
values of the shuffling length ($l_s$). Our goal is to compare the
mutual information $I(X;Y)$ measured in the original SDSS data with
that from the shuffled datasets to test the statistical significance
of $I(X;Y)$ on different length scales.

\subsection{Simulating different morphology-density correlations}

The morphology-density relation is a well known phenomenon which
indicates that environment play a crucial role in deciding galaxy
morphology. We would like to test whether mutual information $I(X;Y)$
can capture the strength of morphology-density relation in the galaxy
distribution. We construct a set of SDSS mock datacubes from a semi
analytic galaxy catalogues as discussed in Section 2.2.

We compute the local number density at the location of each galaxies
using $k^{th}$ nearest neighbour method \citep{casertano85}.  We find
the distance to the the $k^{th}$ nearest neighbour to each galaxy. The
local number density around a galaxy is estimated as,
\begin{eqnarray}
n_k = \frac{k-1}{V(r_k)}  
\label{eqn:knn}
\end{eqnarray}  
Here $r_k$ is the distance to the $k^{th}$ nearest neighbour and
$V(r_k)=\frac{4}{3}\pi r_k^3$. We have used $k=10$ in this analysis.

Our goal is to test if $I(X;Y)$ can capture the degree and nature of
correlation between environment (X) and morphology (Y). The elliptical
galaxies are known to reside preferentially in denser
environments. Each mock SDSS datacubes from the SAM contains a total
$14558$ galaxies. We would like to assign a morphology to each of
these galaxies. To do so, we first sort the number density at the
locations of galaxies in a descending order. We consider three
different schemes which are as follows,

(i) We randomly label $3387$ galaxies as ellipticals from the top 30\%
high density locations and consider the rest of the $11171$ galaxies
as spirals.

(ii) We randomly label $3387$ galaxies as ellipticals from top 50\%
high density locations and consider the rest of the $11171$ galaxies
as spirals.

(iii) We randomly label $3387$ galaxies as ellipticals irrespective of
their local density and consider the rest of the $11171$ galaxies as
spirals.

The morphology-density relation in case (i) is stronger than case (ii)
and there is no morphology-density relation in case (iii). We would
like to test if mutual information $I(X;Y)$ can correctly capture the
degree of association between environment and morphology in these
distributions.

\subsection{Testing statistical significance of the difference in mutual information with $t$ test}
We use an equal variance $t$-test which can be used when both the
datasets consists of same number of samples or have a similar
variance. We calculate the $t$ score at each length scale using the
following formula,
\begin{eqnarray}
t= \frac{|\bar{X_1}-\bar{X_2}|}{\sigma_s \sqrt{\frac{1}{n_1}+\frac{1}{n_2}}}
\label{eqn:ttest}
\end{eqnarray}
where $\sigma_s =
\sqrt{\frac{(n_1-1)\sigma_1^2+(n_2-1)\sigma_2^2}{n_1+n_2-2}}$,
$\bar{X_1}$ and $\bar{X_1}$ are the average values, $\sigma_1$ and
$\sigma_2$ are the standard deviations, $n_1$ and $n_2$ are the number
of datapoints associated with the two datasets at any given
lengthscale. 

We would like to test the null hypothesis that the average value of
mutual information in the original and randomized or shuffled
distribution at a given lengthscale are not significantly
different. We find that randomizing or shuffling the data always leads
to a reduction in the mutual information between morphology and
environment. We use a one-tailed test with significance level
$\alpha=0.0005$ which corresponds to a confidence level of
$99.9\%$. The degrees of freedom in this test is $(n_1+n_2-2)$.  The
same test is also applied to asses the statistical significance of
$I(X;Y)$ in mock datasets where a morphology-density relation is
introduced in a controlled manner. We compute the $t$ score at each
length scale using \autoref{eqn:ttest} and determine the associated $p$
value to test the statistical significance.

\begin{table*}{}
\caption{This table shows the $t$ score and the associated $p$ value
  at each length scale when we compare the mutual information between
  actual SDSS data and SDSS data with randomized morphological
  information.}
\label{tab:ttest1}
\begin{tabular}{ccc}
\hline
 Grid size  ( $\hmpc$ )  & $t$ score & $p$ value \\
\hline
$12.08 $ &	$13.911$ & $2.26 \times 10^{-11}$\\
$13.18 $ &	$13.417$ & $4.10 \times 10^{-11}$\\
$14.50 $ &	$16.125$ & $1.90 \times 10^{-12}$\\
$16.11 $ &	$15.692$ & $3.02 \times 10^{-12}$\\
$18.12 $ &	$20.853$ & $2.34 \times 10^{-14}$\\
$20.71 $ &	$18.698$ & $1.53 \times 10^{-13}$\\
$24.17 $ &	$20.088$ & $4.46 \times 10^{-14}$\\
$29.00 $ &	$28.934$ & $7.59 \times 10^{-17}$\\
$36.25 $ &	$33.613$ & $5.36 \times 10^{-18}$\\
$48.33 $ &	$30.151$ & $3.67 \times 10^{-17}$\\
$72.50 $ &	$30.736$ & $2.61 \times 10^{-17}$\\
\hline
\end{tabular}
\end{table*}  

\begin{table*}{}
\caption{ This table shows the $t$ score and the associated $p$ value
  at each length scale when we compare the mutual information between
  actual SDSS data and its shuffled realizations for different
  shuffling lengths. The grid size for each $n_s$ is chosen in a such
  a way so that the shuffling length is not equal or an integral
  multiple of the grid size.}
\label{tab:ttest2}
\begin{tabular}{cccccccc}
\hline
 Grid size   & \multicolumn{2}{c}{$n_s = 3$}  & \multicolumn{2}{c}{$n_s = 7$}  & \multicolumn{2}{c}{$n_s = 15$}\\
 ( $\hmpc$ ) & $t$ score & $p$ value & $t$ score & $p$ value & $t$ score & $p$ value 	  \\
\hline
$12.08$ &  - 	  & -					 &$ 2.196$ &$ 2.07\times 10^{- 2}$ &$ 2.029$ &$ 2.88\times 10^{- 2}$\\
$13.18$ & $1.559$ &$ 6.82\times 10^{- 2}$ &$ 3.148$ &$ 2.78\times 10^{- 3}$ &$ 4.097$ &$ 3.38\times 10^{- 4}$\\
$14.50$ & $1.967$ &$ 3.24\times 10^{- 2}$ &$ 2.037$ &$ 2.83\times 10^{- 2}$ &$ 4.324$ &$ 2.04\times 10^{- 4}$\\
$16.11$ &  - 	  & -				     &$ 5.064$ &$ 4.04\times 10^{- 5}$ &$ 9.656$ &$ 7.63\times 10^{- 9}$\\
$18.12$ & $3.794$ &$ 6.64\times 10^{- 4}$ &$ 9.806$ &$ 6.03\times 10^{- 9}$ &$13.765$ &$ 2.69\times 10^{-11}$\\
$20.71$ & $4.928$ &$ 5.43\times 10^{- 5}$ & - 	    & - 				      &$13.762$ &$ 2.70\times 10^{-11}$\\
$24.17$ &  - 	  & - 	 		 	     &$12.536$ &$ 1.24\times 10^{-10}$ &$16.367$ &$ 1.48\times 10^{-12}$\\
$29.00$ &$10.667$ &$ 1.64\times 10^{- 9}$ &$20.061$ &$ 4.57\times 10^{-14}$ &$24.534$ &$ 1.38\times 10^{-15}$\\
$36.25$ &$15.510$ &$ 3.68\times 10^{-12}$ &$23.429$ &$ 3.09\times 10^{-15}$ &$29.184$ &$ 6.52\times 10^{-17}$\\
$48.33$ &  - 	  & -				 	 &$26.795$ &$ 2.94\times 10^{-16}$ &$27.235$ &$ 2.20\times 10^{-16}$\\
$72.50$ &$21.376$ &$ 1.52\times 10^{-14}$ &$27.325$ &$ 2.08\times 10^{-16}$ &$29.234$ &$ 6.33\times 10^{-17}$\\
\hline
\end{tabular}
\end{table*} 

\begin{table*}{}
\caption{This table shows the $t$ score and the associated $p$ value at
  each length scale when we compare the mutual information between
  mock datasets with and without a morphology-density relation.}
\label{tab:ttest3}
\begin{tabular}{ccccc}
\hline
 Grid size   & \multicolumn{2}{c}{ Random selection from top 30\%} &  \multicolumn{2}{c} { Random selection from top 50\%} \\
 ( $\hmpc$ ) & 	$t$ score & $p$ value   & 	$t$ score & $p$ value   \\
\hline
$12.08$ &$75.590$ &$ 5.46 \times 10^{-20}$ &$41.511$ &$ 2.32 \times 10^{-16}$\\
$13.18$ &$80.264$ &$ 2.36 \times 10^{-20}$ &$56.285$ &$ 3.35 \times 10^{-18}$\\
$14.50$ &$84.539$ &$ 1.14 \times 10^{-20}$ &$50.987$ &$ 1.33 \times 10^{-17}$\\
$16.11$ &$75.073$ &$ 6.01 \times 10^{-20}$ &$53.361$ &$ 7.04 \times 10^{-18}$\\
$18.12$ &$61.435$ &$ 9.88 \times 10^{-19}$ &$41.186$ &$ 2.59 \times 10^{-16}$\\
$20.71$ &$52.083$ &$ 9.87 \times 10^{-18}$ &$36.618$ &$ 1.32 \times 10^{-15}$\\
$24.17$ &$36.197$ &$ 1.55 \times 10^{-15}$ &$27.185$ &$ 8.10 \times 10^{-14}$\\
$29.00$ &$31.549$ &$ 1.04 \times 10^{-14}$ &$17.937$ &$ 2.34 \times 10^{-11}$\\
$36.25$ &$27.540$ &$ 6.77 \times 10^{-14}$ &$20.213$ &$ 4.65 \times 10^{-12}$\\
$48.33$ &$26.512$ &$ 1.14 \times 10^{-13}$ &$12.707$ &$ 2.23 \times 10^{- 9}$\\
$72.50$ &$ 8.541$ &$ 3.17 \times 10^{ -7}$ &$ 8.379$ &$ 3.98 \times 10^{- 7}$\\
\hline
\end{tabular}
\end{table*}  

\section{Results}

\subsection{Effects of randomizing the morphological classification}

We show the mutual information $I(X;Y)$ between environment and
morphology as a function of length scale in the SDSS datacube in left
panel of \autoref{fig:Imxy_ran} which shows that the
  morphology of the SDSS galaxies and their large-scale environment
  share a small non-zero mutual information throughout the entire
  length scale. The result for the SDSS datasets with randomly
assigned morphology is also shown in the same panel for a
comparison. This shows that there is a significant reduction in
$I(X;Y)$ at each length scale due to the randomization of
morphological information of the SDSS galaxies.  We find that a finite
non-zero mutual information still persists at each length scale even
after the randomization of morphology. To understand its origin, we
also measure the mutual information measured in the Poisson datacubes
with randomly assigned morphology and show them together in the left
panel of \autoref{fig:Imxy_ran}. Interestingly, we find that the
non-zero mutual information between $X$ and $Y$ in the Poisson
distributions are nearly same as the SDSS datacube with randomly
assigned morphology.

The information entropy $H(X)$ associated with environment at each
length scale $d$ remains unchanged, as the position of each galaxies
in the resulting distribution remains same as the original SDSS
distribution. There would be also no change in the information entropy
$H(Y)$ associated with morphology of the galaxies as the number of
spirals and ellipticals remains the same after the
randomization. However this procedure would change the joint entropy
$H(X,Y)$. The randomization of morphological classification would turn
the joint probability distribution to a product of the two individual
probability distribution i.e. $p(X_i,Y_j)=p(X_i)p(Y_j)$. The adopted
procedure is thus expected to destroy any existing correlations
between environment and morphology and consequently any non-zero
mutual information between environment and morphology should ideally
disappear after the randomization.

However in left panel of \autoref{fig:Imxy_ran}, we find that $I(X;Y)$
does not reduce to zero after the randomization of morphology of the
SDSS galaxies. This residual nonzero mutual information can be
explained by the results obtained from the Poisson datacubes with
randomly assigned morphology. The results show that $I(X;Y)$ in the
Poisson datacubes with randomly assigned morphology and SDSS datacube
with randomly assigned morphology are nearly the same. This suggests
that a part of the measured mutual information arises due to the
finite and discrete nature of the galaxy sample. The origin of this
residual information is thus non-physical in nature and should be
properly taken into account during such analysis.

The reduction in $I(X;Y)$ due to the randomization of morphology
suggests that a part of the measured mutual information $I(X;Y)$ must
have some physical origin. Interestingly, left panel of
\autoref{fig:Imxy_ran} shows that randomization leads to a reduction
in the mutual information at each length scale. We test the
statistical significance of these differences at each length scale
using a $t$ test. We show the $t$ score at each length scale in the
right panel of \autoref{fig:Imxy_ran}. The critical $t$ score at
$99.9\%$ confidence level for $18$ degrees of freedom are also shown
in the same panel. The $t$ score and the associated $p$ value at each
length scale are tabulated in \autoref{tab:ttest1}. We find a strong
evidence against the null hypothesis which suggests that the
differences in the mutual information $I(X;Y)$ in the two
distributions are statistically significant at $99.9\%$ confidence
level for the entire length scales probed. This clearly indicates that
the association between environment and morphology is not limited to
only the local environment but extends to environments on larger
length-scales.

\subsection{Effects of shuffling the spatial distribution of galaxies}

We divide the SDSS datacube into a number of regular subcubes using
different values of $l_s$ as discussed in Section 3.3 and shuffle them
many times to generate a set of shuffled realizations for each
shuffling length. The \autoref{fig:3Dv_shuf} shows the distributions
of ellipticals (brown dots) and spirals (blue dots) in the original
unshuffled SDSS datacube along with one realization of the shuffled
datacubes for each shuffling length. The size of the shuffling units
used to shuffle the data in each case are shown with a red subcube at
the corner of the respective shuffled datacubes. A comparison of the
shuffled datacubes with the original SDSS datacube clearly shows that
the coherent features visible in the actual data on larger length
scales progressively disappears with the increasing shuffling
length. It may be noted that both the measurement of $I(X;Y)$ and
shuffling requires us to divide the datacube into a number of
subcubes. In each case, we choose the shuffling lengths and the grid
sizes so that the shuffling length is not equal or integral multiple
of grid size or vice versa. This must be ensured to avoid any spurious
correlations in $I(X;Y)$.

We compare the mutual information $I(X;Y)$ in the original and
shuffled datasets in the left panel of \autoref{fig:Imxy_shuf}. For
each shuffled datasets we observe a reduction in $I(X;Y)$ at different
length scales. A smaller reduction in $I(X;Y)$ is observed at smaller
length scales whereas a relatively larger reduction in $I(X;Y)$ is
seen on larger length scales.

It may be noted that the morphological information of galaxies remain
intact after shuffling the data. The shuffling procedure keeps the
clustering at scales below $l_s$ nearly identical to the original data
but eliminates all the coherent spatial features in the galaxy
distribution on scales larger than $l_s$. Shuffling is thus expected
to diminish any existing correlations between environment and
morphology. Measuring the mutual information between environment and
morphology in the original SDSS data and its shuffled versions allows
us to address the statistical significance of $I(X;Y)$. The mutual
information is expected to reduce by a greater amount on scales above
the shuffling length $l_s$ because shuffling destroys nearly all the
coherent patterns beyond this length scale. On the other hand, we
expect a relatively smaller reduction in $I(X;Y)$ below the shuffling
length $l_s$. This can be explained by the fact that most of the
coherent features in the galaxy distribution below length scale $l_s$
survive the shuffling procedure. However some of the coherent features
which extend upto $l_s$ but lie across the subcubes would be destroyed
by shuffling. Shuffling may also produce a small number of spatial
features which are the product of pure chance alignments. These random
features are unlikely to introduce any physical correlations between
environment and morphology. A comparison of $I(X;Y)$ between the
original and shuffled data at different length scales for different
shuffling length thus reveal the statistical significance of the
degree of association between environment and morphology on different
length scales.

We find that $I(X;Y)$ decreases monotonically at all length scales
with decreasing shuffling lengths. \autoref{fig:Imxy_shuf} shows that
$I(X;Y)$ for $n_s=15$ or $l_s\sim 10 \hmpc$ still lies above the
values that are expected for an identical Poisson random
distributions. A greater reduction in $I(X;Y)$ on larger length scales
for each shuffling length considered suggests that the mutual
information between environment and morphology is statistically
significant on these length scales. $I(X;Y)$ in actual data and
shuffled data for different shuffling lengths do not differ much on
smallest length scale as the coherent structures on these length
scales are nearly intact in all the shuffled datasets. However when
shuffled with smaller values of $l_s$, greater number of coherent
structures on larger length scales are lost. This explain why
reduction in $I(X;Y)$ increases with decreasing shuffling length. 

We employ a $t$ test to test the statistical significance of the
observed differences in $I(X;Y)$ in original and all shuffled datasets
at different length scales. The $t$ score and the corresponding $p$
value at each length scale are tabulated in \autoref{tab:ttest2}. The
$t$ score for the shuffled datasets for three different shuffling
length are shown as a function of length scale in the right panel of
\autoref{fig:Imxy_shuf}. We find that the differences in $I(X;Y)$ in
the shuffled and unshuffled SDSS data are statistically significant at
$99.9\%$ confidence level at nearly the entire length scale probed.

We find a weak evidence against the null hypothesis for all the
shuffling lengths at smaller length scales. This arises due to the
fact that the coherence between environment and morphology are
retained on smaller scales when the data is shuffled with a comparable
or larger shuffling lengths. However we note that a considerable
reduction in $I(X;Y)$ can occur even below the shuffling length for
$n_s=7$ and $n_s=3$. A subset of the coherent features extending below
the shuffling length may lie across the subcubes used to shuffle the
data. These coherent structures will be destroyed by the shuffling
procedure even when they are smaller than the shuffling length. The
number of such coherent structures which belongs to this particular
group is expected to increase with the size of the subcubes due to
their larger boundary.

The results shown in \autoref{fig:Imxy_shuf} thus indicates that the
association between environment and morphology is certainly not
limited to their local environment but extends throughout the length
scales probed in this analysis.

\subsection{Effects of different morphology-density correlations}

In \autoref{fig:3Dv_mill}, we show the distributions of spirals and
ellipticals in mock SDSS datacubes from SAM. We show one distribution
for each of the simulated morphology-density relations.

We show the mutual information $I(X;Y)$ as a function of length scales
for the three different density-morphology relation in the left panel
of \autoref{fig:Imxy_mill}. When the elliptical are randomly selected
from the entire distribution irrespective of their density then we do
not expect any mutual information between morphology and environment.
The non-zero mutual information in this case is just an outcome of the
finite and discrete nature of the distributions. We find that the
results for this case is identical to that expected for a Poisson
distribution with same ratio of spirals to ellipticals.

However when ellipticals are preferentially selected from denser
regions, the mutual information between morphology and environment
rises above the values that are expected for a Poisson random
distribution. The figure \autoref{fig:Imxy_mill} shows that mutual
information $I(X;Y)$ is significantly higher than Poisson distribution
when galaxies are randomly tagged as elliptical from the top $50\%$
high density positions. We find that the mutual information between
morphology and environment increase further to much higher values when
galaxies are randomly identified as ellipticals from the top $30\%$
high density regions. We note a change in $I(X;Y)$ at all lengthscales
upto $50 \hmpc$. A larger change in $I(X;Y)$ is observed on smaller
length scales whereas the change in $I(X;Y)$ becomes gradually smaller
on larger length scales. This indicates that the morphology-density
relations simulated here, become weaker on larger length scales.

We use a $t$ test to asses the statistical significance of the
differences in $I(X;Y)$ in the mock datasets with and without a
morphology-density relation. We tabulate the $t$ score and the
corresponding $p$ value at each length scale for the two mock datasets
in \autoref{tab:ttest3}. In the right panel of
\autoref{fig:Imxy_mill}, we show the $t$ score as a function of length
scales in two mock datasets with different morphology-density
relation. The results suggest that a statistically significant
difference ($99.9\%$ confidence level) exists between the datasets
with and without a morphology-density relation. Interestingly, these
differences persist throughout the entire length scale probed in the
analysis. This indicates that the correlation between environment and
morphology is not limited to the local environment but extends to
larger length scales.

The morphology-density relations considered here are too simple in
nature. In this experiment, we find that the mutual information
between morphology and environment decreases monotonically with
increasing length scales. Contrary to this, the SDSS observations show
that mutual information initially decreases with increasing length
scales and nearly plateaus out at larger length scales. The schemes
used for the morphology-density relation in this experiment are not
realistic in nature. But they clearly shows that mutual information
can effectively capture the degree of association between morphology
and environment and such a relation may extend upto larger length
scales.

\section{Conclusions}

In the present work, we aim to test the statistical significance of
mutual information between morphology of a galaxy and its environment.
The morphology-density relation is a well known phenomenon which has
been observed in the galaxy distribution. The relation suggests that
the ellipticals are preferentially found in denser regions of galaxy
distribution whereas spirals are sporadically distributed across the
fields. It is important to understand the role of environment in
galaxy formation and evolution. The local density at the location of a
galaxy is very often used to characterize its environment. It is
believed that the environmental dependence of galaxy properties can be
mostly explained by the local density alone. The mutual information
between environment and morphology for SDSS galaxies has been studied
by \citet{pandey17} where they find that a non-zero mutual information
between morphology and environment persists throughout the entire
length scale probed. They show that the mutual information between
environments on different length scales may introduce such
correlations between environment and morphology observed on larger
length scales. We would like to critically examine the statistical
significance of the observed non-zero mutual information between
morphology and environment on different length scales. We propose
three different methods to asses the statistical significance of
mutual information. These methods also help us to understand the
relative importance of environment on different length scales in
deciding the morphology of galaxies.

Three different tests are carried out in the present analysis. In the
first case, we randomize the morphological information about the SDSS
galaxies without affecting their spatial distribution. In the second
case, we shuffle the spatial distribution of the SDSS galaxies without
affecting their morphological classification. Both these tests show
that the mutual information between morphology and environment are
statistically significant at $99.9\%$ confidence level throughout the
entire length scales probed in this analysis. We find that a small
non-zero mutual information can be observed even in a random
distribution without any existing physical correlations between
environment and morphology. This non-zero value originates from the
finite and discrete nature of the distribution. Interestingly, the
mutual information between environment and morphology in the SDSS
datacube is significantly larger than the randomized datasets
throughout the entire length scales probed. Shuffling the SDSS
datacube also affect the mutual information between environment and
morphology in a statistically significant way at nearly the entire
length scales considered. This suggests that the association between
morphology and environment continues upto a larger length scales and
these correlations must have a physical origin. In a third test, we
construct a set of mock SDSS datacubes from the semi analytic galaxy
catalogue where we assign morphology to the simulated galaxies based
on the density at their locations. We vary the strength of the
simulated morphology-density relation and measure mutual information
between environment and morphology in each case. Our results suggest
that mutual information effectively capture the degree of association
between environment and morphology in these mock datasets.

  We extend our analysis to dark matter halo sample from
  Millennium simulation (see Appendix A) where we investigate if the
  angular momentum of dark matter halos display any large-scale
  correlations at fixed halo mass. The analysis shows that
  statistically significant correlations are observed only for the
  halos in the mass range $\sim 10^{11}-10^{12} M_{\odot}$. The
  assembly bias is known to be more pronounced at low masses ($\sim
  10^{12} M_{\odot}$) and the observed correlations could be a
  signature of assembly bias. But we could not confirm this due to a
  wider variation of halo mass in this mass range. Choosing a narrower
  range around this halo mass does not provide us sufficient number of
  dark matter halos within the specific volume required for the
  present analysis. The present analysis also suggests that the
  observed large-scale correlations between morphology and environment
  is small but statistically more significant than that observed
  between the angular momentum of dark matter halos and their
  environment.

    Besides the halo assembly bias, the rich baryonic
    physics may also play an important role, which allow much more
    complicated interactions between galaxies and their
    environment. The effects of local density on morphology of
    galaxies is understood in terms of various types of galaxy
    interactions, ram pressure stripping and quenching of star
    formation. These processes may play a dominant role in shaping the
    morphology of a galaxy. However they may not be the only factors
    which decides the morphology of a galaxy. The presence of
    large-scale coherent features like filaments, sheets and voids may
    induce large-scale correlations between the observed galaxy
    properties and their environment. Further studies may reveal if
    any new physical processes are required to explain such
    large-scale correlations. In any case, we need to understand the
    physical origin of such correlations and if required, incorporate
    them in the models of galaxy formation. Most studies employ
    correlation functions to study the assembly bias. Here, we
    speculate that the information theoretic framework presented in
    this paper, might serve as a more sensitive probe of galaxy
    assembly bias than traditional correlation functions.

Every statistical measure have their pros and cons. One particular
drawback of mutual information is that it does not tell us the
direction of the relation between two random variables i.e. the
measured mutual information does not provide us the simple information
that the ellipticals and spirals are preferentially distributed in
high density and low density regions respectively. But the mutual
information reliably captures the degree of association between any
two random variables irrespective of the nature of their
relationship. So in the present context, mutual information can be an
effective and powerful tool to quantify the degree of influence that
environment imparts on morphology across different length scales. The
amplitude of mutual information quantify the strength of correlation
between morphology and environment on different length scales. It also
helps us to probe the length scales upto which the morphology of a
galaxy is sensitive to its environment.

One can also study the mutual information between environment and any
other galaxy property to understand the influence of environment on
that property at various length scales. The relative influence of
environment on different galaxy properties on any given length scale
may provide useful inputs for the galaxy formation models. Finally we
note that mutual information between environment and a galaxy property
is a powerful and effective tool which can be used successfully for
the future studies of large-scale environmental dependence of galaxy
properties.

\section{Data availability}
The data underlying this article are available in
https://skyserver.sdss.org/casjobs/ and
https://www.mpa.mpa-garching.mpg.de/millennium/. The datasets were
derived from sources in the public domain: https://www.sdss.org/,
http://zoo1.galaxyzoo.org and
https://www.mpa.mpa-garching.mpg.de/millennium/.

\section{ACKNOWLEDGEMENT}
The authors thank an anonymous reviewer for useful comments and
suggestions which helped us to improve the draft. SS would like to
thank UGC, Government of India for providing financial support through
a Rajiv Gandhi National Fellowship. BP would like to acknowledge
financial support from the SERB, DST, Government of India through the
project CRG/2019/001110. BP would also like to acknowledge IUCAA, Pune
for providing support through associateship programme.
  
The authors would like to thank the SDSS team and Galaxy Zoo team for
making the data public. Funding for the Sloan Digital Sky Survey IV
has been provided by the Alfred P. Sloan Foundation, the
U.S. Department of Energy Office of Science, and the Participating
Institutions. SDSS-IV acknowledges support and resources from the
Center for High-Performance Computing at the University of Utah. The
SDSS web site is www.sdss.org.

SDSS-IV is managed by the Astrophysical Research Consortium for the
Participating Institutions of the SDSS Collaboration including the
Brazilian Participation Group, the Carnegie Institution for Science,
Carnegie Mellon University, the Chilean Participation Group, the
French Participation Group, Harvard-Smithsonian Center for
Astrophysics, Instituto de Astrof\'isica de Canarias, The Johns
Hopkins University, Kavli Institute for the Physics and Mathematics of
the Universe (IPMU) / University of Tokyo, the Korean Participation
Group, Lawrence Berkeley National Laboratory, Leibniz Institut f\"ur
Astrophysik Potsdam (AIP), Max-Planck-Institut f\"ur Astronomie (MPIA
Heidelberg), Max-Planck-Institut f\"ur Astrophysik (MPA Garching),
Max-Planck-Institut f\"ur Extraterrestrische Physik (MPE), National
Astronomical Observatories of China, New Mexico State University, New
York University, University of Notre Dame, Observat\'ario Nacional /
MCTI, The Ohio State University, Pennsylvania State University,
Shanghai Astronomical Observatory, United Kingdom Participation Group,
Universidad Nacional Aut\'onoma de M\'exico, University of Arizona,
University of Colorado Boulder, University of Oxford, University of
Portsmouth, University of Utah, University of Virginia, University of
Washington, University of Wisconsin, Vanderbilt University, and Yale
University.

The Millennium Simulation data bases \citep{lemson} used in this paper
and the web application providing online access to them were
constructed as part of the activities of the German Astrophysical
Virtual Observatory.

\appendix

\begin{figure*}
\resizebox{9.5 cm}{!}{\rotatebox{0}{\includegraphics{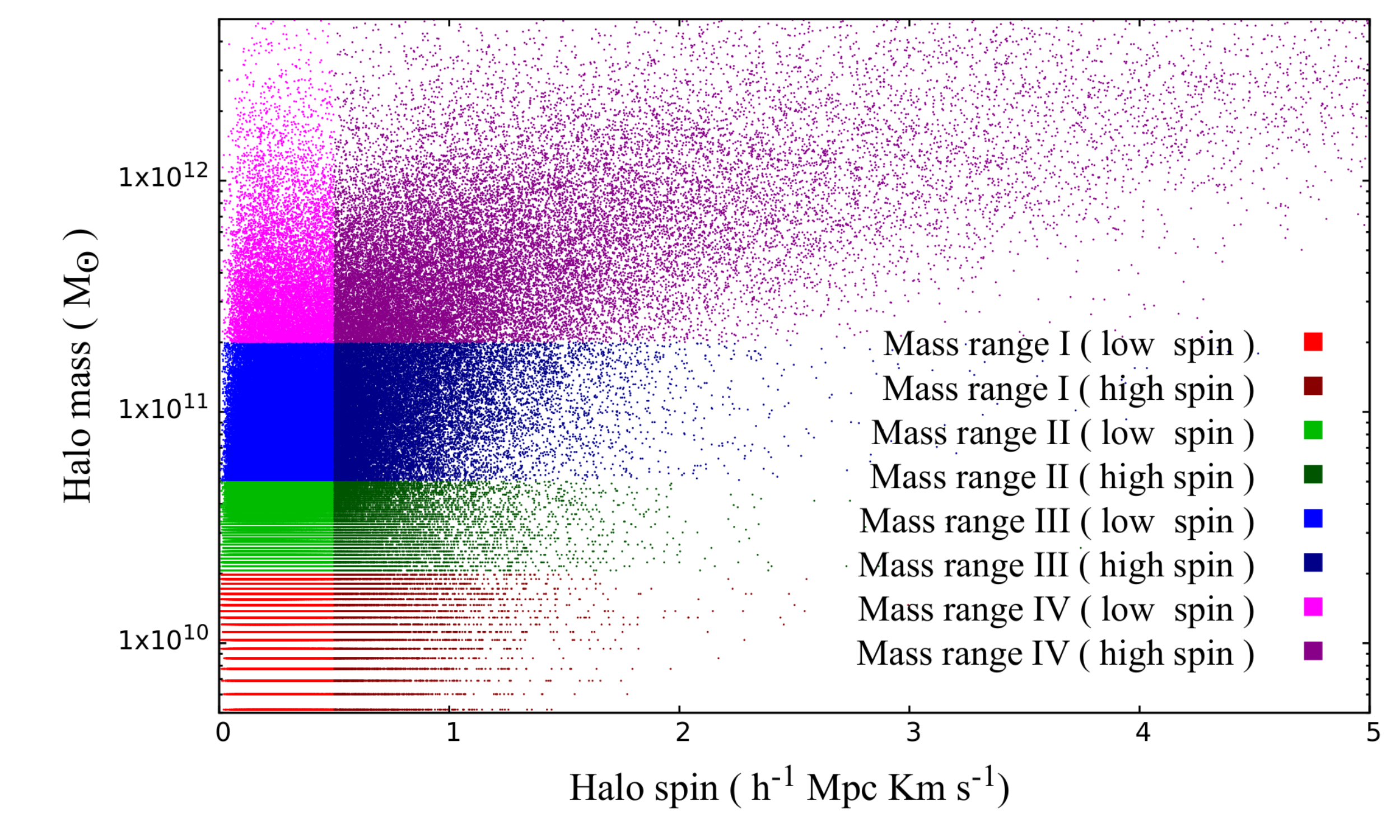}}} \hspace {1cm}
\resizebox{6 cm}{!}{\rotatebox{0}{\includegraphics{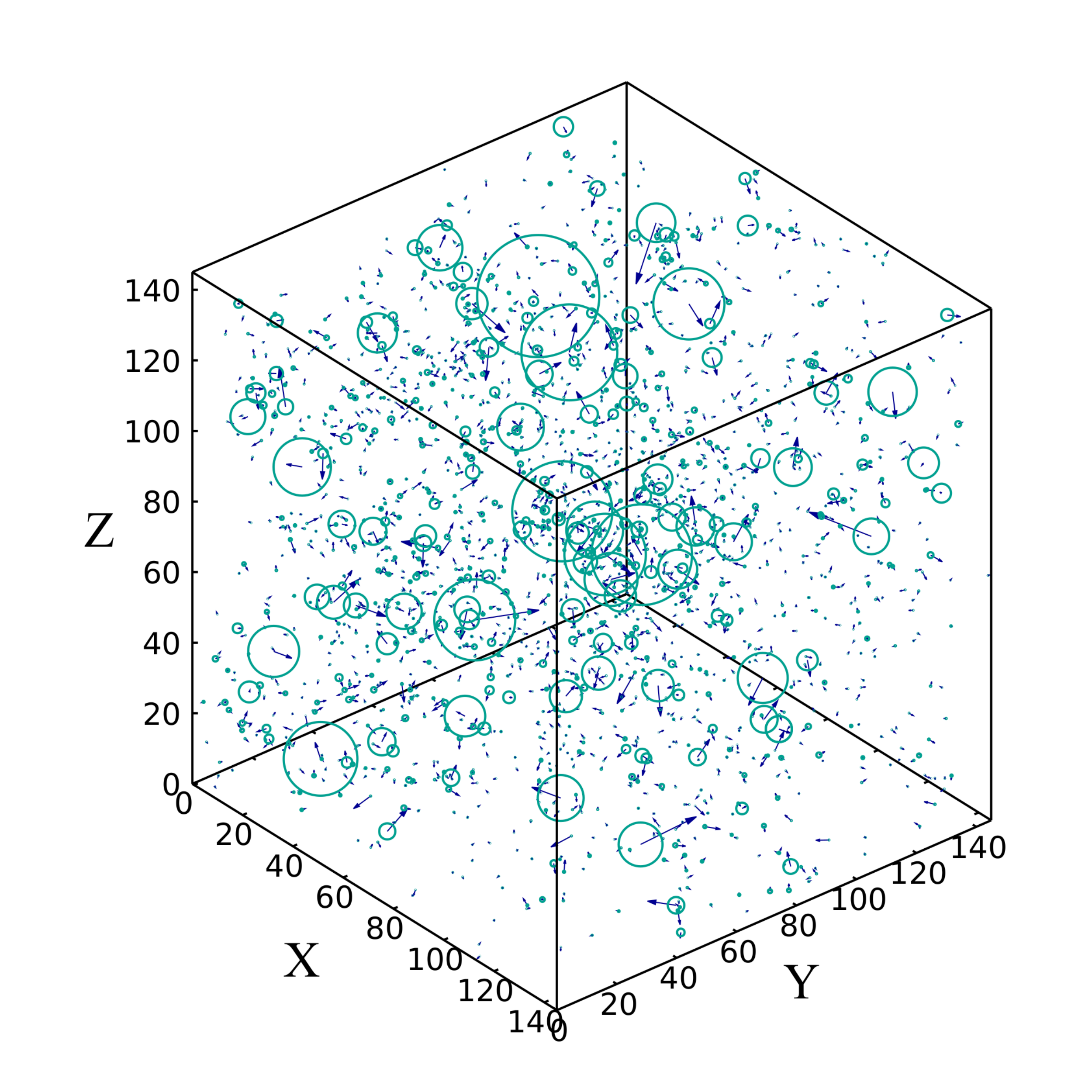}}}
\caption{The left panel of the figure shows the
    distribution of halo spin in the four mass bins. The right panel
    of the figure shows the spatial distribution of 2000 dark matter
    halos randomly selected from the entire population.  In this
    panel, each green circle represents a dark matter halo and the
    blue arrows attached to each circle represent the angular
    momentum vector associated with the halo. The radii of the circles
    are proportional to the masses and the lengths of the arrows are
    proportional to the magnitudes of the angular momentum of the dark
    matter halos. This is shown as a visual representation of the
    distribution of dark matter halos and their angular momentum. We
    have used only the spatial distribution of the halo centers in our
    analysis.}
  \label{fig:spinVmass}
\end{figure*}

\begin{figure*} 
\resizebox{16 cm}{!}{\rotatebox{0}{\includegraphics{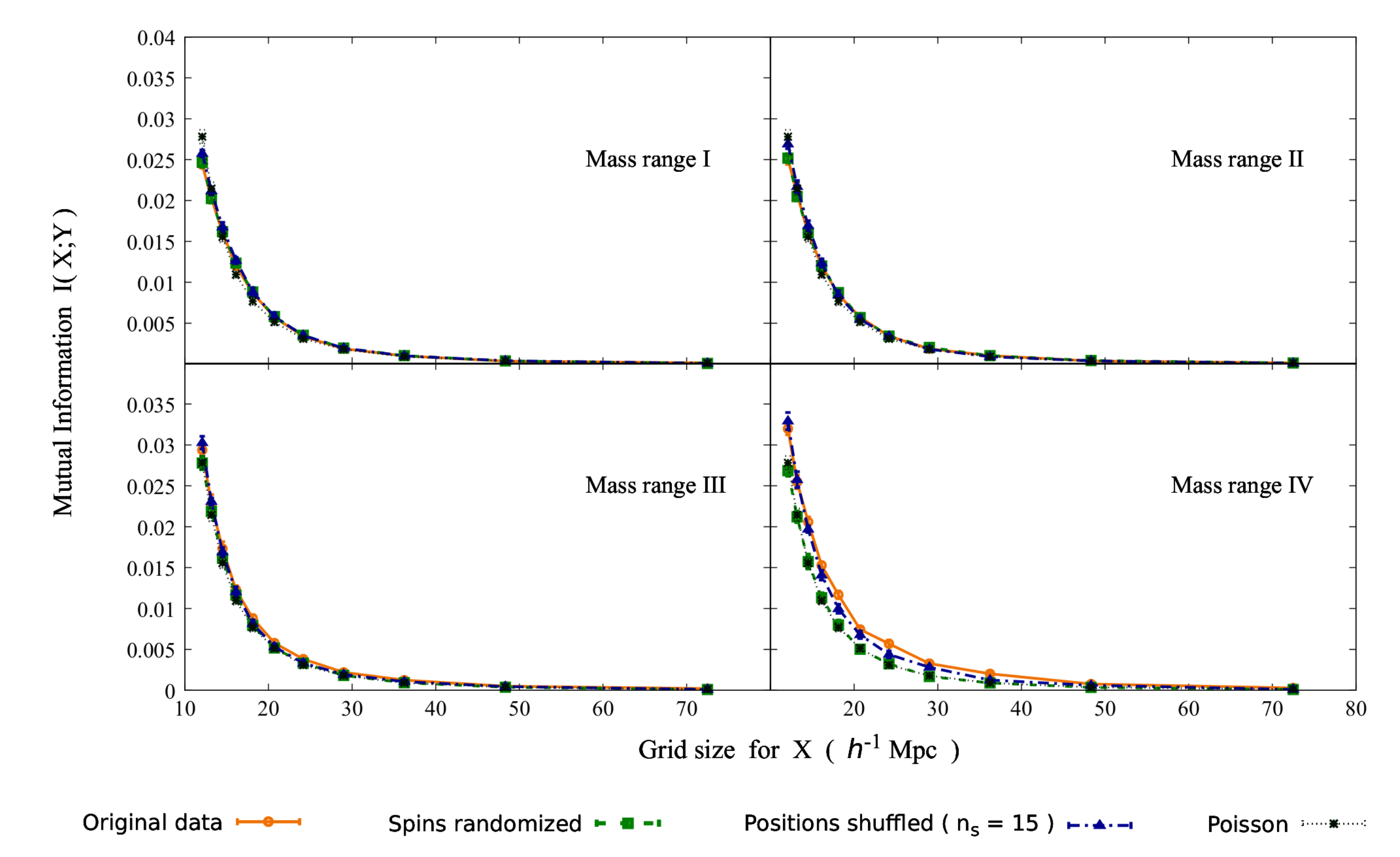}}}
\caption{The different panels of this figure show the mutual
  information between the environment of a dark matter halo and its
  spin angular momentum for $4$ different mass bins. The results after
  randomization of the halo spins and shuffling of the spatial
  distributions ($n_s=15$) are also shown together in each panel. We
  also show the results for mock Poisson samples with randomly
  assigned spin angular momentum in each case. The $1-\sigma$
  errorbars are obtained from 10 subsamples analyzed in each mass
  bin.}
  \label{fig:mh_mr}
\end{figure*}

\begin{figure*} 
\resizebox{16 cm}{!}{\rotatebox{0}{\includegraphics{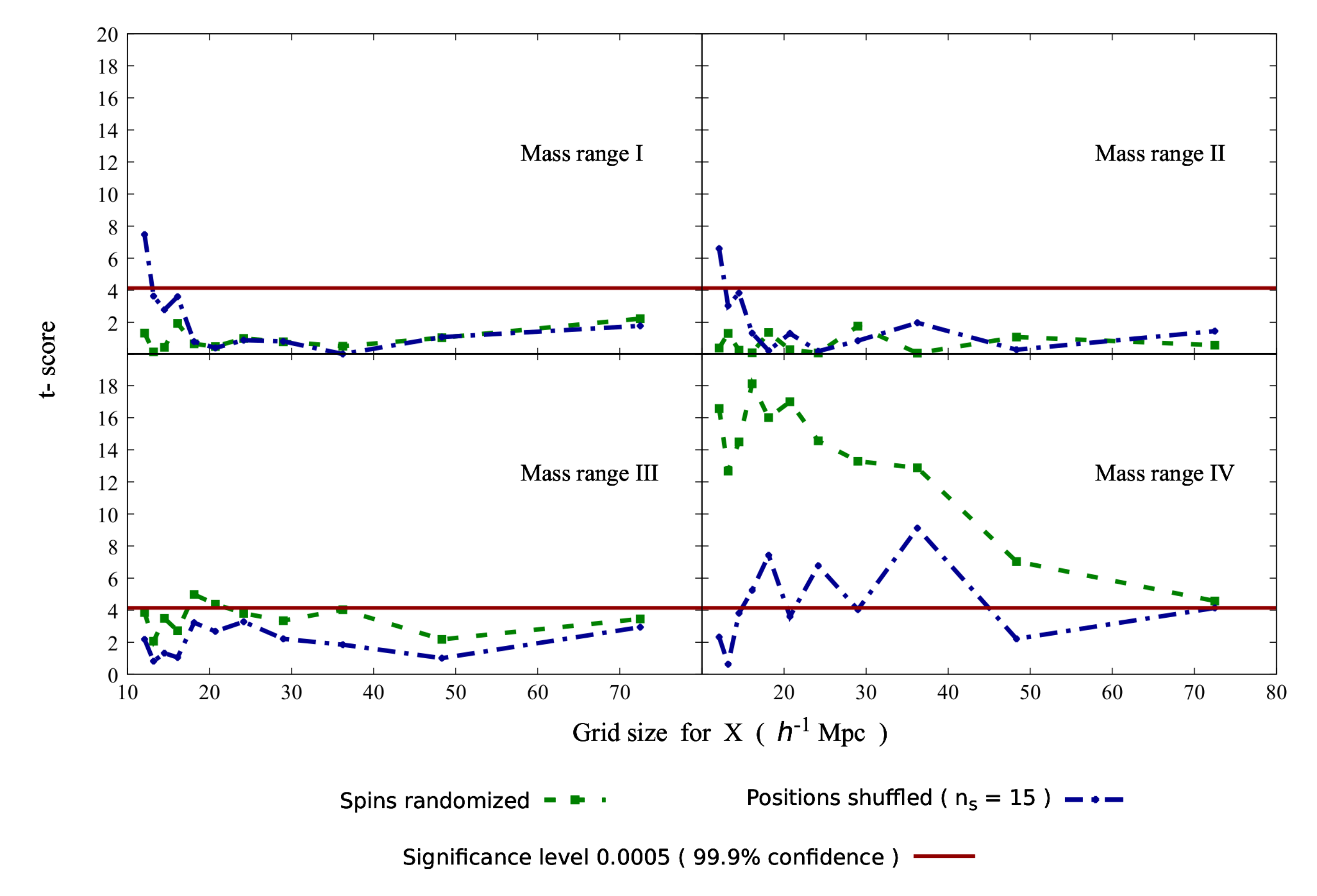}}}
\caption{The different panels of this figure show the t score at
  different length scales, for the randomized and shuffled
  distributions in four different halo mass ranges.}
  \label{fig:tt_mr}
\end{figure*}

\begin{table*}{}
\caption{This table shows the four bins which is used to classify the dark matter halos.}
\label{tab:mass_bins}
\begin{tabular}{ll}
\hline
Mass range   I ( in $M_{\odot}$ unit )  & : \, $ 5\times 10^{9} \le M \le 2\times 10^{10} $\\
Mass range  II ( in $M_{\odot}$ unit )  & : \, $ 2\times 10^{10} \le M \le 5\times 10^{10} $\\
Mass range III ( in $M_{\odot}$ unit )  & : \, $ 5\times 10^{10} \le M \le 2\times 10^{11} $\\
Mass range IV  ( in $M_{\odot}$ unit )  & : \, $ 2\times 10^{11} \le M \le 5\times 10^{12} $\\
\hline
\end{tabular}
\end{table*}  

\section{Mutual information between the environment and angular momentum of dark matter halos}
In this section, we would like to extend our analysis to dark matter
halos using the Millennium halo catalogue. We download the data from
{\it MPAHaloTrees} table of Millennium database using a SQL query. We
retrieve the virial mass, angular momentum (in $h^{-1}$Mpc Km s$^{-1}$
units ) and 3 dimensional position and velocities of all the dark
matter halos. We map the spatial distribution of the halos from real
space to redshift space and extract a cubic region of size $145
\hmpc$. These halos are then divided into four different bins
depending on their masses, which are defined in
\autoref{tab:mass_bins}. We construct $10$ mock samples corresponding
to each mass bin. $14558$ halos are randomly selected for $10$ times
for each mass bin. Each mock sample thus contains $14558$ dark matter
halos distributed within a cubic region of size $145 \hmpc$.

Our goal is to perform an analysis with the dark matter halos to test
the large-scale environmental dependence of halo spins. The analysis
is analogous to the study we carried out for the SDSS galaxies. We use
the magnitude of angular momentum of the dark matter halos to classify
them into two different groups. The distribution of the angular
momentum ($ \omega $) of the dark matter halos in different mass bins
are shown in the left panel of \autoref{fig:spinVmass}. Only the dark
matter halos with $ 0 < \omega \leq 5 $ are considered in this
analysis. We use a critical value $\omega_c = 0.5$ to divide the halos
in two different classes. The choice of $\omega_c$ is somewhat
arbitrary. We choose this value to have significant number of halos in
both the high and low angular momentum states. Halos with $ \omega
\leq \omega_c $ are termed as low spin halos whereas the ones with
$\omega_c < \omega \leq 5$ are labelled as high spin halos.

  In order to randomize the angular momentum of the
  halos, we randomly select pairs of halos and swap their spin tags
  without altering their positions. The number of low and high spin
  halos remains unchanged after such randomization. For each mock
  sample, we randomly identify $100 \times 14558$ pairs and
  interchange their spin states. Any correlation between the
  environment of a halo and its spin is expected to be destroyed by
  this operation.

  We then shuffle the spatial distribution of the dark
  matter halos keeping their angular momentum unchanged. We consider
  the smallest shuffling length ($n_s=15$) used for the analysis with
  SDSS galaxies. Higher values of shuffling lengths would be necessary
  only if there is a significant change in the mutual information
  introduced by this shuffling length.

  We calculate the mutual information for the randomized
  and shuffled distributions of dark matter halos in each mass range.
  
  In \autoref{fig:mh_mr}, we show the mutual
  information between the environment of a halo and its angular
  momentum as a function of length scale for four different mass
  bins. We find a small nonzero mutual information between environment
  and angular momentum which respectively extend upto $30 \hmpc$ and
  $50 \hmpc$ in the first two and last two mass bins. We compare these
  mutual information with those obtained for the randomized and
  shuffled distributions in each mass bin. We find that randomizing the
  spins and shuffling the data do not introduce any noticeable change
  in the measured mutual information between angular momentum and
  environment in the first two halo mass bins. The fact that the
  actual mutual information coincides with that from randomized,
  shuffled and Poisson data at all scales suggests that these non-zero
  mutual information do not have any physical origin. They purely
  arise due to finite and discrete nature of the
  distributions. However, we observe that randomization and shuffling
  change the mutual information in the mass bins III and IV. We asses
  the statistical significance of the differences in each mass bin
  using $t$- test. The results are shown in different panels of
  \autoref{fig:tt_mr}. Clearly the low mass bins (bin I and II) do not
  show a statistically significant change in the mutual information
  over nearly the entire length scale. The statistical significance of
  the differences gradually increases with halo mass (bin III and IV)
  as can be seen in the two bottom panels of \autoref{fig:tt_mr}. A
  statistically significant difference ($99.9\%$ confidence) is
  observed over most of the length scales for the halos in mass range
  IV.

  This analysis shows that for smaller mass halos, there
  are no clear association between the angular momentum of dark matter
  halos and their large-scale environment. A statistically significant
  correlation between these variables are observed only for the
  relatively more massive dark matter halos in mass bin IV. The
  assembly bias is expected to be more significant for halo mass $\sim
  10^{12} M_{\odot}$ which are included in bin IV. So these
  correlations, may in principle originate from assembly
  bias. Alternatively, they may be a manifestation of a wider
  variation of halo mass in bin IV. Unfortunately, we could not verify
  this due to lack of sufficient number of halos within the chosen
  volume when a narrow mass range is opted for bin IV. We also repeat
  our analysis with different values of $\omega_c$ and recover the
  same trend. Finally, we note that the correlations between
  morphology of SDSS galaxies and their large-scale environment are
  more pronounced than the associations detected between angular
  momentum of dark matter halos and their environment.

\bsp	
\label{lastpage}
\end{document}